\documentclass{article} 
\usepackage{iclr2026_conference,times}

\usepackage{hyperref}
\usepackage{multirow}
\usepackage{url}
\usepackage{enumitem}
\usepackage{caption}
\usepackage{amsmath}
\usepackage{amssymb}
\usepackage{graphicx}
\usepackage{xspace}
\usepackage{xcolor}
\usepackage{tcolorbox}
\usepackage{wrapfig}

\newcommand{\sys}{\textsc{FSA}\xspace}

\newcommand{\ryan}[1]{\begingroup\color{black}#1\endgroup}

\title{\sys: An Alternative Efficient Implementation of Native Sparse Attention Kernel}

\author{Ran Yan$^{1*}$, Youhe Jiang$^{1*}$, Zhuoming Chen$^{2}$, Haohui Mai$^{1}$, Beidi Chen$^{2}$, Binhang Yuan$^{1}$  \\
1. The Hong Kong University of Science and Technology \\
2. Carnegie Mellon University\\
\texttt{ryanaf@connect.ust.hk, youhejiang@gmail.com} \\ \texttt{\{zhuominc, beidic\}@andrew.cmu.edu, \{haohui, biyuan\}@ust.hk}
}

\iclrfinalcopy 
\begin{document}


\maketitle

\begingroup
\renewcommand{\thefootnote}{}
\footnotetext{$^*$ represents equal contribution. Correspondence to Binhang Yuan.}
\endgroup

\begin{abstract}
Recent advances in sparse attention mechanisms have demonstrated strong potential for reducing the computational cost of long-context training and inference in large language models (LLMs). Native Sparse Attention (NSA), one state-of-the-art approach, introduces natively trainable, hardware-aligned sparse attention that delivers substantial system-level performance boosts while maintaining accuracy comparable to full attention. However, the kernel implementation of NSA forces a loop order that is only efficient with a relatively large number of query heads in each Grouped Query Attention (GQA) group, whereas existing LLMs widely adopt a much smaller number of query heads in each GQA group --- such an inconsistency significantly limits the applicability of this sparse algorithmic advance. In this work, we propose \textbf{\underline{F}lash \underline{S}parse \underline{A}ttention (\sys)}, an alternative kernel implementation that enables efficient NSA computation across a wide range of popular LLMs with a varied, smaller number of heads in each GQA group on modern GPUs. Compared to vanilla NSA kernel implementation, our empirical evaluation demonstrates that \sys achieves (\underline{i}) up to 3.5$\times$ and on average 1.6$\times$ kernel-level latency reduction, (\underline{ii}) up to 1.25$\times$ and 1.09$\times$ on average end-to-end training speedup on state-of-the-art LLMs, and (\underline{iii}) up to 1.36$\times$ and 1.11$\times$ on average for prefill-phase speedup in LLM generative inference. The source code is open-sourced and publicly available at \url{https://github.com/Relaxed-System-Lab/Flash-Sparse-Attention}.
\end{abstract}

\section{Introduction}
\label{sec:intro}


Large Language Models (LLMs) with long context windows~\citep{gpt4o,claude3,young2024yi,dubey2024llama} face prohibitive computational costs due to the full attention mechanism’s quadratic time and memory complexity. As sequence length increases, attention computation becomes a critical bottleneck --- for instance, attention can account for $70–80\%$ of total decoding latency at a 64k token context~\citep{yuan2025native}. In extreme cases, processing a 1 million-token prompt with an 8B model can take up to 30 minutes on a single GPU~\citep{jiang2024minference}. These observations underscore the urgent need for more efficient attention mechanisms in long-context LLM training and inference. A recent promising direction is to exploit sparse attention, whereby the query of each token only interacts with a subset of key and value, dramatically reducing the computation load and HBM I/O volumes. However, implementing efficient sparse attention at scale is non-trivial --- In fact, the challenge of implementing high-performance kernels has become a major obstacle to deploying state-of-the-art sparse attention techniques in practice. In this paper, we want to explore: \textit{Can we design and implement an efficient sparse attention kernel for a wide range of current LLMs to fully unleash the potential of this algorithmic advance over modern GPUs?}

Addressing the above question is crucial because adopting sparse attention in long-context LLMs could mitigate the quadratic cost and enable new applications~\citep{xu2025128k, chen2024core, acharya2024star, wang2024beyond}. By leveraging the inherent sparsity of attention patterns, one can significantly cut down computation and memory overhead. Among such methods, one promising example is Natively Sparse Attention (NSA)~\citep{yuan2025native}, a recently proposed sparse attention framework, which organizes keys/values into blocks and processes them via three parallel attention modules --- compressed coarse-grained tokens, selected fine-grained tokens, and sliding local windows. By learning which tokens to compress or drop, NSA achieves long-context efficiency without a predefined pattern, making it a natural choice for long-context LLM training. 

Nevertheless, implementing an efficient sparse attention kernel, i.e, NSA, is challenging. The core difficulty lies in implementing the sparse mechanism in NSA (i.e., computing attention score based on selectively retained fine-grained tokens), where the query of each token needs to dynamically select a different set of keys and values. Such computation results in irregular HBM access patterns on modern GPUs, where each query processes distinct selected keys/values, potentially requiring unnecessary padding for query tiles before executing warp-/warpgroup- level matrix multiply-and-accumulate instructions (e.g., \texttt{wmma} or \texttt{wgmma}), and leading to the underutilization of tensor cores.

This scattered access pattern conflicts with the GPU hardware-efficient design principle: GPUs achieve their peak mathematical throughput when the warps execute dense (no-padded) matrix multiply and accumulation instructions. Thus, current sparse attention \ryan{implementations} fail to translate the theoretical floating-point operations (FLOPs) reduction into wall-clock speedups. 


Vanilla NSA kernel implements a two-level loop: In the outer loop, NSA kernel loads one token and batches query attention heads that share the same key and value heads; in the inner loop, NSA kernel loads selected KV block iteratively and performs attention computation. This strategy reaches kernel efficiency only when each Grouped Query Attention (GQA)~\citep{ainslie2023gqa} group has sufficient number of query heads, so that no-padding is required to execute PTX instructions (e.g., \texttt{wmma} or \texttt{wgmma}) on modern GPUs.\footnote{Concretely, performance is downgraded due to hardware requirements on matrix shapes for warp-/warpgroup- level matrix multiply-and-accumulate instructions (e.g., \texttt{wmma} or \texttt{wgmma})~\citep{ptx_warp_level_matrix}, where each dimension of a matrix tile must be larger than specified value (e.g., at least 8 on Hopper GPUs).} However, such an assumption may not hold for a wide range of popular LLMs so that the original NSA kernel efficiency could drop considerably. With an insufficient number of query heads in each GQA group, batching query heads is inefficient to satisfy this hardware requirement. Thus, the original NSA kernel implementation must pad query attention heads to meet instruction requirements, resulting in unnecessary data loading and computations. 







To resolve this issue, we propose \sys, which implements optimized kernels efficient for NSA under various GQA group settings. We make the following concrete contributions:

\begin{itemize}[topsep=5pt, leftmargin=*]
\vspace{-0.5em}
\item \textbf{\underline{Contribution 1:}} We propose an alternative implementation for the NSA kernel, which exchanges the two-level loop order in NSA implementation --- \sys loops over KV blocks in the outer loop and loops over query tokens in the inner loop to accelerate this system bottleneck. Since the number of query tokens that attend to a given KV block is usually much larger than the hardware required value, \sys introduces no padding, significantly reducing unnecessary kernel memory access and FLOPs, thereby facilitating faster token selection kernel execution.

\item \textbf{\underline{Contribution 2:}} We analyze the trade-off between vanilla NSA and \sys implementation in terms of kernel efficiency and memory accessing paradigm, which illustrates the effective design and implementation of \sys. To maximize the performance benefits of \sys kernel design, we implement dedicated optimizations for query token memory access, which is accessed in the inner loop of \sys kernel, and employ separate optimized kernels for attention result reduction.


\item \textbf{\underline{Contribution 3:}} We conduct empirical studies to compare \sys with vanilla NSA and full attention. Concretely, we benchmark kernel execution latencies, end-to-end training and inference prefill phase latencies for state-of-the-art LLMs. Compared to NSA, results show that \sys delivers (\underline{i}) up to 3.5$\times$ and on average 1.6$\times$ kernel-level latency reduction, (\underline{ii}) up to 1.25$\times$ and 1.09$\times$ on average end-to-end training speedup, and (\underline{iii}) up to 1.36$\times$ and 1.11$\times$ on average inference prefill-phase speedup. Compared to full attention, the performance boost is further amplified.
\vspace{-0.5em}
\end{itemize}

\section{Preliminaries and Related Work}
\label{sec:preliminaries}

\subsection{GPU Kernel Implementation}

\textbf{Parallelization in modern GPUs.} Modern GPUs utilize massive threads to execute kernels concurrently. Optimized kernel implementations typically employ two-level parallelism: (\underline{i}) Thread block-level parallelism: Optimized implementations partition input matrices into multiple tiles, assign them to thread blocks, and execute computations for each thread block in parallel. Common paradigm within a single thread block follows three key steps: Load matrix tiles into the GPU's shared memory; perform computations using the loaded tiles; and store computed results to the output tensor. (\underline{ii}) Warp-level parallelism: Within each thread block, optimized kernels further partition assigned matrix tiles to multiple warps --- each containing 32 threads on NVIDIA GPUs~\citep{nvidia_cuda_best_practices} --- to enable fine-grained parallel execution. Warp-level parallelism maximizes hardware efficiency through coalesced memory access and implicit synchronization within warps.

\textbf{Efficient kernel implementation.} Modern GPU architectures impose strict requirements on the shapes of matrix tiles used in low-level computations. Specifically, PTX warp-level matrix multiply-accumulate instructions~\citep{ptx_warp_level_matrix} require that for matrix multiplication $C=AB$, where $A\in \mathbb{R}^{m\times k}$ and $B\in \mathbb{R}^{k\times n}$, the dimensions $m$, $n$, and $k$ must satisfy minimum size requirements for single-warp processing. On NVIDIA Hopper GPUs, $m$, $n$, $k$ must be at least 8. To achieve higher efficiency, a thread block typically utilizes multiple warps for sufficient warp-level parallelism. Additionally, modern GPUs perform optimally with coalesced and contiguous data loading and storing; non-contiguous memory access leads to a lower L2 cache hit rate, thereby reducing effective memory bandwidth and degrading overall kernel efficiency.

\subsection{Attention Mechanisms}
\label{sec:attn}

\textbf{Full attention.} Full attention with causality~\citep{vaswani2017attention, ainslie2023gqa}---where each query token attends to all previous KV tokens---is standard in LLM training and inference. Formally, given sequence length $N$, query/key head dimension $d_K$, value head dimension $d_V$, $h$ query heads, and $h_K$ KV heads, attention computation involves query/key/value tensor $\mathbf{Q}\in \mathbb{R}^{N\times d_K \times h}$, $ \mathbf{K}\in \mathbb{R}^{N\times d_K \times h_K}$, $ \mathbf{V}\in \mathbb{R}^{N\times d_V \times h_K}$. For $j$-th ($j\in \{0,1,...,h-1\}$) query head, $\lfloor j / h_K \rfloor$-th (ranging from 0 to $h_K$-1) key and value head, denote involved matrices as $\mathbf{Q}^j, \mathbf{K}^{\lfloor j / h_K \rfloor}\in\mathbb{R}^{N\times d_K},\mathbf{V}^{\lfloor j / h_K \rfloor} \in\mathbb{R}^{N\times d_V}$. Full attention computation can be formalized as:

\begin{small}
\begin{equation}
\begin{split}
\mathbf{O}^{j} = \text{Softmax}\left(\frac{\mathbf{Q}^j(\mathbf{K}^{\lfloor j / h_K \rfloor})^T }{\sqrt{d_K}}\right)\mathbf{V}^{\lfloor j / h_K \rfloor}
\end{split}
\end{equation}
\end{small}%

On the system side, recent research~\citep{dao2023flashattention, kwon2023efficient} has optimized full attention from various perspectives. Notably, Flash Attention~\citep{dao2023flashattention} optimizes full attention with a two-level loop: Each thread block loads a block of query tokens and, while KV tokens remain, iteratively processes a block of KV tokens and accumulates intermediate results with online softmax~\citep{milakov2018online}. Results are finally written to the output tensor. This design minimizes redundant memory accesses for query and output tensors, thereby reducing attention execution latency.

\textbf{Sparse attention.} Recent efforts in sparse attention algorithms ~\citep{yuan2025native, lu2025moba, Lee2023SEA, Tay2020SparseSinkhornAttention, Zhao2019ExplicitSparseTransformer, tang2024quest, xiaoduoattention, zhu2024sampleattention, laiflexprefill, xu2025xattention, zhang2023h2o} and system side optimizations efforts~\citep{zhang2024sageattention, zhang2024sageattention2, zhang2025sageattention3} represent an emerging trend aimed at reducing attention computation costs in long-context LLM training and inference, where standard attention performs poorly due to its quadratic complexity with respect to sequence length. The most notable efforts in sparse attention include Native Sparse Attention (NSA)~\citep{yuan2025native}. Formally, in NSA, for $j$-th query head, each query token $\mathbf{q}^j_t \in \mathbb{R}^{1 \times d_K}, t\in \{0,1,...,N-1\}$ attends to $\tilde{N} \ll N$ KV tokens via three attention mechanisms $c \in \mathcal{C}$, where $\mathcal{C} = \{\text{cmp}, \text{sel}, \text{win}\}$, representing compression, selection, and sliding window for keys and values. We denote KV tokens as $\mathbf{\tilde{K}}_c^{\lfloor j/h_K  \rfloor}\in \mathbb{R}^{\tilde{N} \times d_K}, \mathbf{\tilde{V}}_c^{\lfloor j/h_K  \rfloor} \in \mathbb{R}^{\tilde{N} \times d_V}$, which contains $\lfloor j/h_K\rfloor$-th KV head and a subset of KV tokens of attention mechanism $c$. Given trainable gating scores $\tau^c_t \in [0, 1]$ for three attention modules, NSA combines the three attention mechanisms as follows:

\begin{small}
\begin{equation}
\begin{split}
\mathbf{o}^j_t = \sum_{c\in \mathcal{C}} \tau^c_t \cdot \text{Softmax}\left(\frac{\mathbf{q}^j_t \mathbf{(\tilde{K}}_c^{\lfloor j / h_K \rfloor})^T}{\sqrt{d_K}}\right)\tilde{\mathbf{V}}_c^{\lfloor j / h_K \rfloor}
\end{split}
\end{equation}
\end{small}%

Notably, the NSA kernel that selectively retains fine-grained tokens is a major system bottleneck across three attention mechanisms. This point is validated in \S \ref{sec:case_abl_study}. The NSA kernel allows each query token across query heads that share the same KV heads to attend to distinct $T$ KV blocks, each with $B_K$ contiguous KV tokens. Distinct KV block selection imposes challenges on effectively batching query tokens and performing computation with KV blocks within one thread block. Therefore, it is crucial to optimize the batching strategy for efficient NSA kernel execution.

\section{Flash Sparse Attention}
\label{sec:fsa}

We present \sys design and compare with vanilla NSA (\S\ref{sec:fsa-design}), then introduce \sys implementation and optimizations (\S\ref{sec:fsa-impl}). Finally, we provide a thorough analysis of \sys performance (\S\ref{sec:fsa-analysis}).

\subsection{\sys Kernel Design}
\label{sec:fsa-design}

\begin{figure}[t!]
    \centering
    \includegraphics[width=\linewidth]{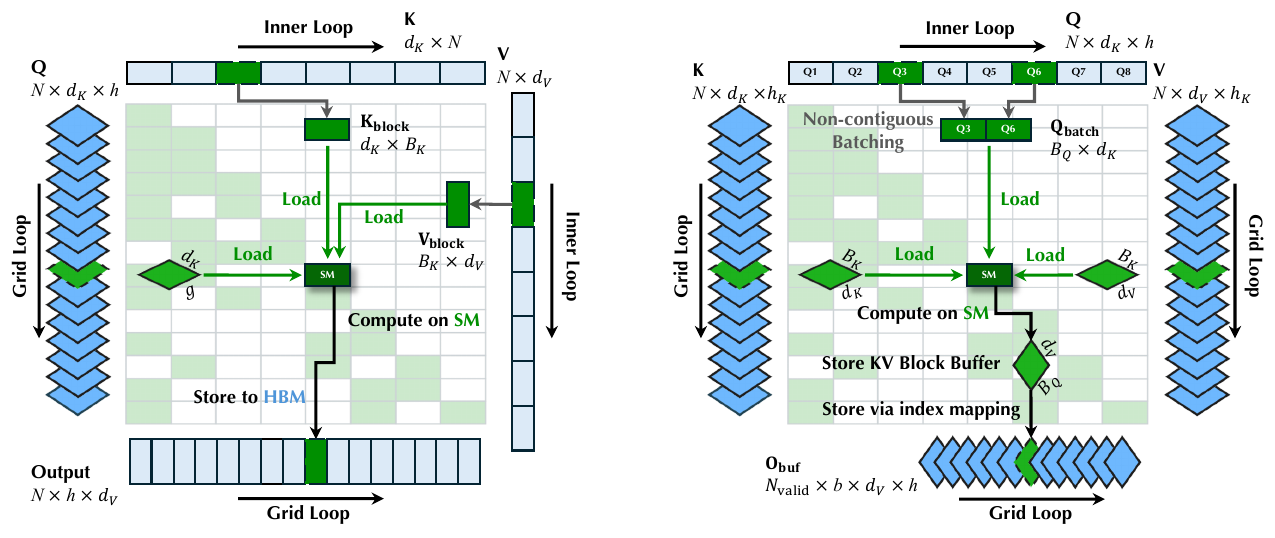}
    \caption{\underline{Left}: Illustration of NSA kernel~\citep{yuan2025native}, which iterates query tokens in the outer loop, and processes KV blocks in the inner loop. \underline{Right}: Illustration of \sys kernel, which alternatively iterate KV blocks in the outer loop, and processes query tokens in the inner loop --- partial attention results are stored in output buffer $\mathbf{O}_{\text{buf}}$ for accumulation (see \S\ref{sec:fsa-impl} for more details).}
    \label{fig:kernel_design}
\end{figure}

An efficient sparse attention kernel must translate theoretical FLOPs reduction into concrete savings in memory access and computation during GPU execution. Vanilla NSA kernel is insufficient in achieving this goal. As illustrated in Figure \ref{fig:kernel_design} (left), NSA kernel processes query tokens one by one in the outer loop and KV blocks in the inner loop, while batching query heads. However, if the number of query heads is insufficient, this method requires padding to meet the hardware's matrix multiplication shape requirements, leading to wasteful memory access and computation. 

To achieve higher kernel efficiency, \sys exchanges NSA kernel loop order and processes query heads one by one, looping over KV blocks in the outer loop and batches of query tokens in the inner loop. Since the number of such tokens is typically large enough to meet hardware requirements, this strategy requires no padding and eliminates the overhead of processing padded data.

However, due to inversion of kernel loop order, \sys encounters new challenges:

\begin{itemize}[topsep=5pt, leftmargin=*]
\vspace{-0.5em}

\item \textbf{Non-contiguous memory access for query batches.} Due to the sparse nature of NSA token selection, for one KV block, only a subset of total query tokens is involved for attention computation, and query token indices are typically non-contiguous. When processing query tokens in \sys inner loop, it is critical to minimize the negative impact of non-contiguous memory access.

\item \textbf{Online softmax statistics and attention results accumulation.} Online softmax and attention results reduction for each query token across distinct KV blocks adds another layer of complexity. In the NSA token selection logic, computing the final output for a query token requires accumulating partial attention results from its distinct selected KV blocks. Since the NSA kernel's outer loop iterates over query tokens, this accumulation process can be handled within one thread block. In contrast, \sys's inverted loop order means that partial results for a single query are computed across different thread blocks, each processing a different KV block. This design necessitates a proper management strategy for accumulating attention results distributed across thread blocks.
\vspace{-0.5em}
\end{itemize}


\subsection{\sys Kernel Implementation and Optimization}
\label{sec:fsa-impl}

To implement an efficient \sys kernel, we employ an optimized token selection kernel that minimizes the negative impact of non-contiguous memory access. Additionally, an online softmax and reduction kernel are designed to efficiently handle online softmax and attention result reduction.

\textbf{\sys token selection kernel.} \textit{\sys mitigates the impact of non-contiguous memory access by employing index tensors to orchestrate data movement.} During forward pass, as illustrated in Figure \ref{fig:kernel_design} (right), each thread block in \sys kernel is assigned a single (Query Head, KV Block) pair. The KV block is loaded from main memory once per thread block. The kernel then iterates through batches of non-contiguous query tokens, which are loaded and stored using index tensors $\mathcal{I}_i$ and $\mathcal{O}_i$ for $i \in \{1, 2, ..., b\}$, where $b$ is the total number of KV blocks. These index tensors are pre-computed from the NSA sparse selection tensor $\mathbf{T} \in \mathbb{R}^{h_K \times N \times T}$, which records selected KV block indices for each query token. Due to the sparse nature of token selection, each KV block is attended by a subset of $N$ query tokens. Consequently, index tensor $\mathcal{I}_i$, which contains query token indices attending to current KV block, typically holds fewer than $N$ valid indices, i.e., $N_{\text{valid}} = |\mathcal{I}_i| \leq N$. To minimize the impact of non-contiguous memory access, a thread block terminates early once it has processed all valid query indices in $\mathcal{I}_i$, avoiding further memory access or computation. Concurrently, index mapping tensor $\mathcal{O}_i$ facilitates contiguous storage of intermediate results. Note that outputs from \sys token selection kernel are not final attention scores; they are partial results that are reduced for each query across different KV blocks in a separate reduction kernel, which we introduce next. In the backward pass, \sys kernel follows a similar logic, loading query tokens non-contiguously to compute gradients and storing intermediate gradients to buffers. The primary difference is that index tensors $\mathcal{I}_i$ and $\mathcal{O}_i$, computed during the forward pass, are retrieved from cache.


\textit{\sys handles query attention results and gradients reduction in separate kernels.} In forward pass, \sys parallel computation of attention scores --- where a single query token's results are reduced across multiple KV blocks --- requires a careful implementation of online softmax and reduction logic to ensure numerical correctness. In the backward pass, a similar reduction challenge exists for gradients of query tokens. FSA achieves efficient and correct accumulation in two kernels:

\textbf{\sys reduction kernel.}
Since a query's attention scores or gradients are computed across multiple thread blocks (each processing a different KV block in \sys token selection kernel), direct reduction into the output tensor in \sys kernel necessitates atomic additions~\citep{atomic_add} to prevent race conditions. Given the prohibitive overhead of atomic operations, \sys decouples computation from accumulation. It adopts a two-stage process: 

\begin{itemize}[topsep=5pt, leftmargin=*]
\vspace{-0.5em}

\item (\underline{i}): \sys token selection kernel (see Figure \ref{fig:kernel_design} (right)) computes partial query attention results or gradients without reduction with online softmax and writes them to an intermediate buffer.

\item (\underline{ii}): A dedicated reduction kernel efficiently accumulates these partial results into a final output tensor with online softmax scaling, which we introduce next. 
\vspace{-0.5em}
\end{itemize}
This two-stage arrangement effectively eliminates atomic operations and achieves efficient attention result accumulation. However, HBM memory overhead is increased due to intermediate buffers. To minimize memory overhead, we allocate a buffer sized only for $N_{\text{valid}}$ query tokens relevant to each KV block, rather than for all $N$ tokens. Index mapping tensor $\mathcal{O}_i$ facilitates contiguous I/O into this compact buffer, thereby avoiding the significant overhead of allocating a full-sized buffer for each KV block. We present a detailed analysis of \sys buffer HBM memory overhead in Appendix \ref{app:memory}.

\textbf{\sys online softmax kernel}. In the forward pass, to ensure numerical correctness, \sys needs to include online softmax statistics in two aspects:

\begin{itemize}[topsep=5pt, leftmargin=*]
\vspace{-0.5em}

\item (\underline{i}): In the \sys token selection kernel, computation results between each query token and key block must be scaled with \underline{\textit{historical}} running maximum~\citep{milakov2018online}). 

\item (\underline{ii}): In the reduction kernel, partial attention outputs of query tokens regarding selected KV blocks stored in the output buffer must be scaled with online softmax statistics. Additionally, final output for a query token must be scaled with log-sum exponentials~\citep{milakov2018online}.
\vspace{-0.5em}
\end{itemize}

Computing online softmax statistics within the \sys token selection kernel produces incorrect attention results. When multiple thread blocks process the same query token, each block computes only \underline{\textit{partial}} statistics, leading to incorrect maximum values and attention outputs. To address this challenge, \sys introduces a separate online softmax kernel that pre-computes online softmax statistics using query and key tensor $\mathbf{Q}$ and key tensor $\mathbf{K}$ and stores them in a buffer. 

\begin{wrapfigure}{r}{0.5\linewidth}
    \centering
    \vspace{-2em}
    \includegraphics[width=\linewidth]{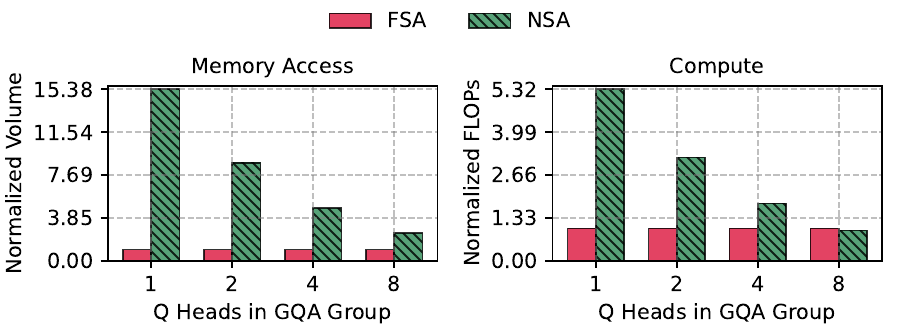}
    \vspace{-1em}
    \caption{Comparison on memory access and FLOPs, block size is 64, top-k is 16. \sys's memory volume or FLOPs are normalized to 1.}
    \vspace{-2em}
    \label{fig:ablation}
    
\end{wrapfigure}

\subsection{\sys Performance Analysis}
\label{sec:fsa-analysis}

We analyze \sys performance by answering two critical questions regarding \sys and NSA kernel performance:

\textbf{Question 1:} \textit{Do additional auxiliary kernels like online softmax and reduction implemented in \sys incur additional memory access and computation overhead?}

To answer this question, we conduct detailed memory footprint and computation load analysis and derive the following theorem:

\textbf{Theorem:} \textit{Across popular GQA group settings, where each GQA group contains $g\in \{1,2,4,8 \}$ query heads, aggregate memory access volume and FLOPs of \sys token selection, online softmax, and reduction kernel are lower than vanilla NSA kernel.} Comparisons are presented in Figure \ref{fig:ablation}. Additional memory access introduced by auxiliary kernels, i.e., \sys online softmax and reduction kernels, remains manageable, falling significantly below memory access wasted on padded data in the original NSA kernel (see more details in Appendix \ref{app:memory}).


\begin{wrapfigure}{r}{0.5\linewidth}
    \centering
    \vspace{-2em}
    \includegraphics[width=\linewidth]{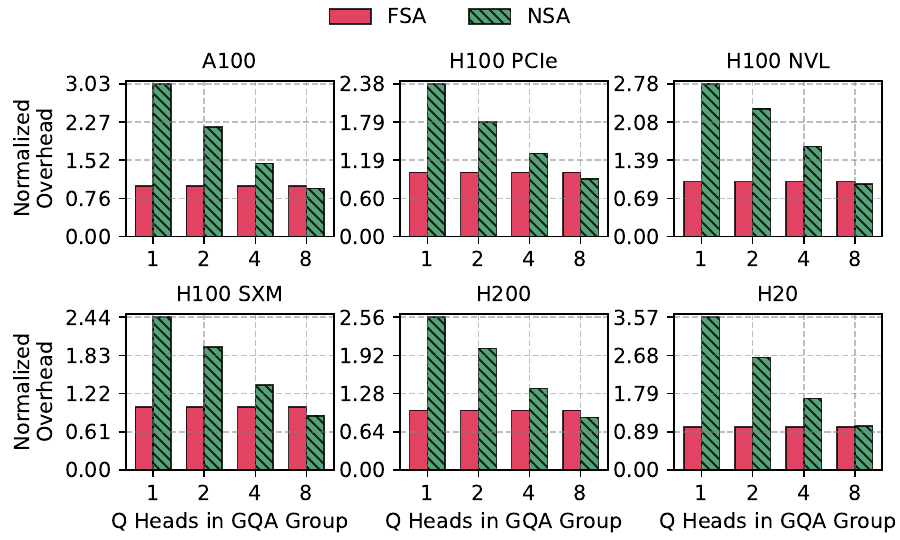}
    \vspace{-1em}
    \caption{Real-time profiling results of the \sys and NSA kernel execution overhead across different GPUs, under block size $B_K=64$, and top-k value $T=16$. \sys latency is normalized to 1.}
    \vspace{-2em}
    \label{fig:profile}
\end{wrapfigure}

\textbf{Question 2:} \textit{Since \sys introduces non-contiguous memory access on loading query tokens and requires additional auxiliary kernels, is \sys generally applicable across various GPU types, and does it consistently provide performance improvements over NSA kernels?}

To answer this question, we conduct a group of micro-benchmarks and enumerate the following analysis of empirical results:

\textbf{Empirical analysis:} \textit{Profiling results (shown in Figure \ref{fig:profile}) across various GPU types and GQA group settings confirm superior performance of \sys.} Optimized \sys outperforms vanilla NSA across popular GPU architectures and GQA group settings, despite being compromised by non-contiguous memory access, and reducing attention results in a separate kernel. When each GQA group contains fewer than 8 query heads, \sys usually demonstrates superior performance to NSA. These empirical results demonstrate that \sys kernel's performance gains from overall reduced unnecessary memory access and FLOPs more than compensate for the overhead of non-contiguous memory access and executing multiple kernels.

\section{Evaluation}
\label{sec:eval}
This section presents a comprehensive evaluation of \sys across various NSA configurations. We aim to investigate the following research questions:
\begin{itemize}[topsep=5pt, leftmargin=*]
\vspace{-0.5em}
\item \textit{Q1: What is the kernel-level performance of \sys compared with NSA and full attention across diverse NSA algorithmic configurations?}
\item \textit{Q2: What is the impact of \sys on end-to-end training and inference performance in practice?}
\item \textit{Q3: What is the breakdown performance of \sys, and how effective is each part of \sys?}
\vspace{-0.5em}
\end{itemize}

\subsection{Experimental Setup}
\label{subsec:setup}

\textbf{Experimental setups.} We use two GPU types for evaluations: NVIDIA H20 GPUs~\citep{h20}, which provide 148 TFLOPS tensor core computational power and 4 TB/s memory bandwidth; and NVIDIA H200 GPUs~\citep{h200specs}, which deliver 989 TFLOPS tensor core computational power and 4.8 TB/s memory bandwidth. For end-to-end training and inference evaluations, GPUs are interconnected via NVLink, providing 450 GB/s inter-GPU bandwidth. \ryan{In our evaluations, we use BF16 for training and FP16 for inference.}

\textbf{Baselines.} We compare \sys with two baselines:
\begin{itemize}[topsep=5pt, leftmargin=*]
\vspace{-0.5em}
\item \textbf{NSA (Native Sparse Attention)~\citep{yuan2025native}.} Our primary baseline is vanilla NSA implementation, which introduces natively hardware-aligned trainable sparse attention. NSA maintains algorithmic performance comparable to full attention while substantially reducing computational complexity. We utilize Triton-based NSA kernel~\citep{nativesparseattention} for evaluation.

\item \textbf{Full attention (Flash Attention) \citep{dao2023flashattention}.} Due to limited hardware resource utilization, theoretical FLOPs reductions achieved by NSA or \sys may not translate to proportional performance gains. Therefore, the full attention baseline (with causality), which has no sparsity constraints, is essential to demonstrate the practical effectiveness of both NSA and \sys. We utilize an efficient Triton-based Flash Attention kernel\citep{tritonattention} for fair comparison.
\vspace{-0.5em}

\end{itemize}

\begin{figure}[t]
    \centering
    \includegraphics[width=\linewidth]{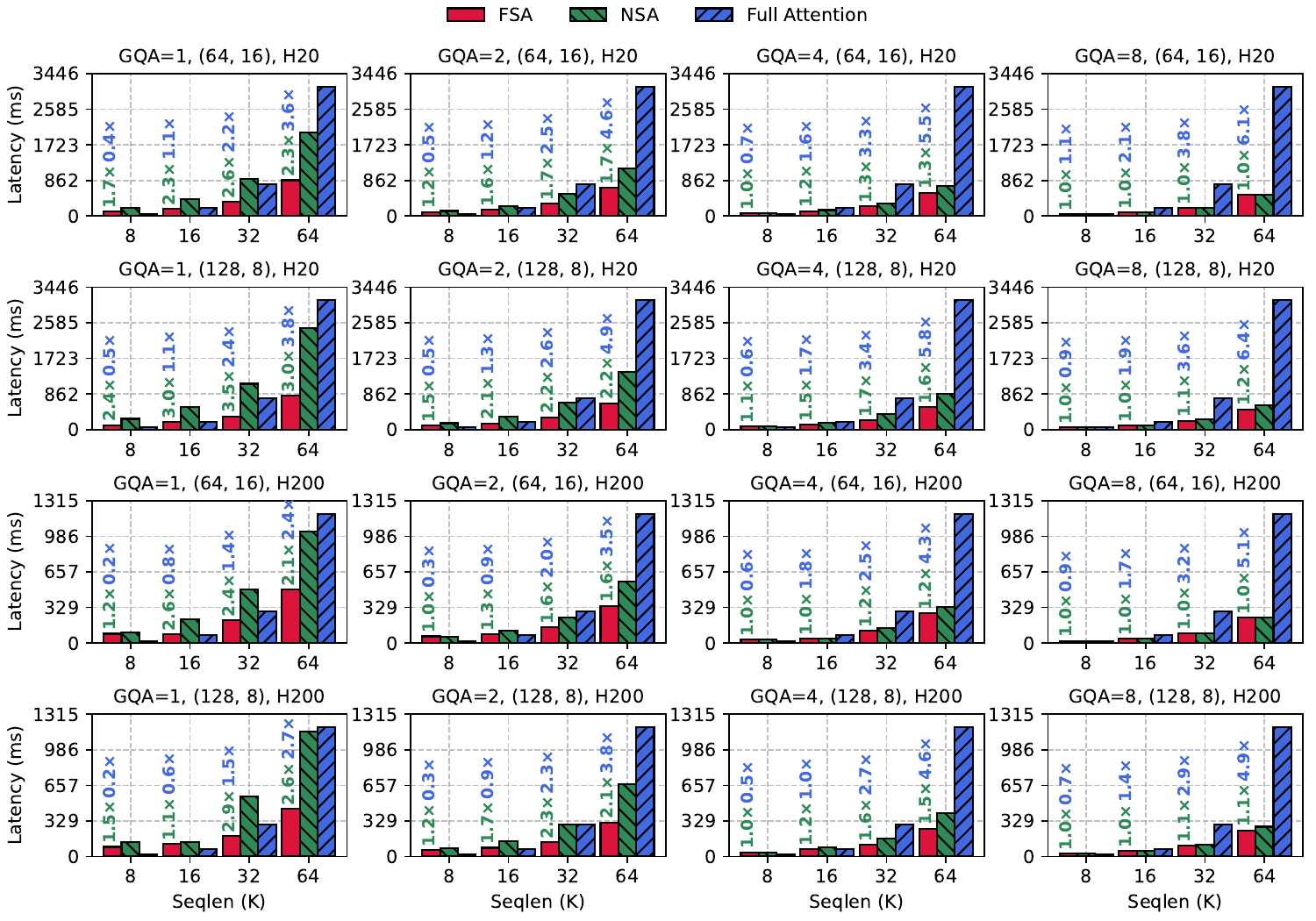}
\caption{Performance comparison of Triton-based \sys, NSA, and full attention (enabled by Flash Attention) kernels under block sizes and top-k values of ($B_K$, $T$) equals to ($64$, $16$) and ($128$, $8$).}
    \label{fig:kernel-e2e}
\end{figure}

\textbf{Experimental configurations.} To ensure comprehensive evaluation, we systematically test \sys and two baselines under varying NSA configurations: (\underline{i}) GQA settings $g \in \{1, 2, 4, 8\}$, where $g$ is number of query heads in one GQA group; (\underline{ii}) NSA hyperparameter block size $B_K$ and top-k $T$ combinations of $(B_K, T) \in \{(64,16), (128,8)\}$; and (\underline{iii}) sequence lengths of $\{8\text{K}, 16\text{K}, 32\text{K}, 64\text{K}\}$ tokens.\footnote{More experiments on ultra long sequence lengths, i.e., 128K and 256K sequence lengths, are presented in Appendix \ref{app:ultra-long}.} For end-to-end training and inference evaluations, we evaluate performance using Llama3-8B~\citep{dubey2024llama}, Qwen3-14B~\citep{yang2025qwen3}, and Qwen2.5-32B~\citep{team2024qwen2} with sequence lengths of 32K and 64K. When the entire model is too large to fit on a single GPU for training, we use pipeline parallelism~\citep{shoeybi2019megatron} to distribute model across multiple GPUs.

\textbf{Evaluation metrics.} Following established practices in prior research~\citep{yuan2025native,lu2025moba,dao2023flashattention}, we employ two metrics to evaluate system efficiency: (\underline{i}) Kernel execution latency, which measures computational time required for attention operations, and (\underline{ii}) training and inference latency, which measures end-to-end time required to process a single batch of data during model training and inference. These metrics directly assess \sys's computational efficiency.

\subsection{\sys Kernel Benchmarking Results (Q1)}
\label{sec:kernelbench}

\textbf{\sys kernel performance.} We evaluate the kernel performance of \sys across both H20 and H200 GPUs under various configurations. In this section, we evaluate \sys on one single GPU, while we present distributed evaluations of \sys in Appendix \ref{app:dist}. As shown in Figure \ref{fig:kernel-e2e}, the evaluation results demonstrate that \sys outperforms both NSA and full attention across most of the tested scenarios:
\begin{itemize}[topsep=5pt, leftmargin=*]
\vspace{-0.5em}
\item \textbf{Comparison with NSA.} \sys outperforms NSA with significantly lowered memory access volume and FLOPs in NSA token selection module, despite introducing non-contiguous memory access and auxiliary kernels (see details in \S\ref{sec:fsa}). \sys achieves up to 3.5$\times$ speedup and on average 1.8$\times$ lower kernel latency on H20 GPUs, and up to 2.9$\times$ speedup and on average 1.4$\times$ lower kernel latency on H200 GPUs compared to NSA. Performance gap between \sys and NSA widens with smaller GQA group settings ($g \in \{1,2\}$) and longer sequence lengths (32K and 64K tokens), with peak performance improvement of 3.5$\times$ observed at $g=1$ (one query head in one GQA group) and sequence length of 32K tokens. Furthermore, \sys maintains consistent performance improvements across different NSA algorithmic configurations, e.g., where $(B_K, T)=(64,16)$ and $(B_K, T)=(128,8)$, demonstrating robust efficiency gains across diverse parameter settings.

\item \textbf{Comparison with full attention.} For long sequences, \sys outperforms full attention with an efficient NSA algorithm and even more efficient token selection. \sys achieves up to 6.4$\times$ speedup and on average 2.4$\times$ lower kernel latency on H20 GPUs, and up to 4.9$\times$ speedup and on average 2.3$\times$ lower kernel latency on H200 GPUs compared to full attention. Performance gap between \sys and full attention increases dramatically with a larger number of query heads in each GQA group, with the most substantial improvement of 6.4$\times$ observed at $g=8$ (8 query heads in one GQA group) and sequence length of 64K tokens. Similarly, \sys maintains superior efficiency across $(B_K, T)\in \{(64,16), (128, 8)\}$ settings, demonstrating consistent and substantial performance advantages over full attention. On the other hand, vanilla NSA lags behind full attention in many tested cases, even with its sparse attention mechanism. For example, when the sequence length is 32K, one GQA group contains one query head, NSA consistently falls short of full attention, while \sys demonstrates superior performance than full attention.
\vspace{-0.5em}
\end{itemize}

\begin{figure}[t!]
    \centering
    \begin{minipage}[b]{0.48\textwidth}
        \centering
        \includegraphics[width=\linewidth]{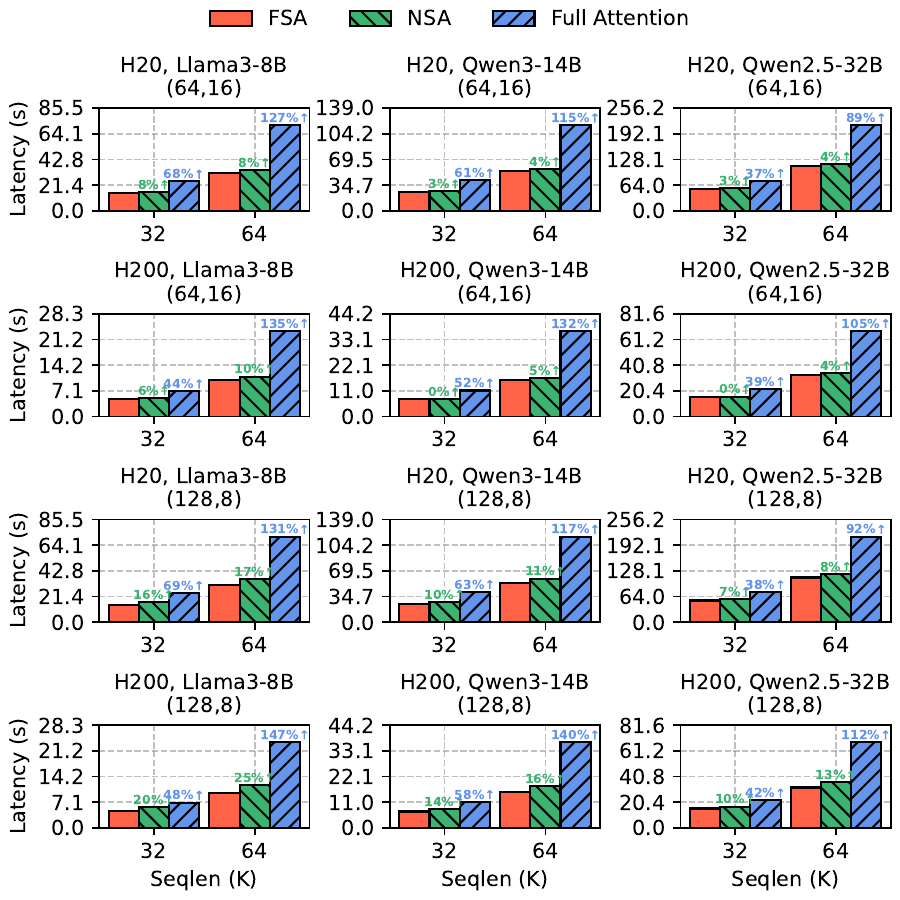}
        \caption{End-to-end training latency of \sys, NSA, full attention.}
        \label{fig:e2e}
    \end{minipage}
    \hfill
    \begin{minipage}[b]{0.48\textwidth}
        \centering
        \includegraphics[width=\linewidth]{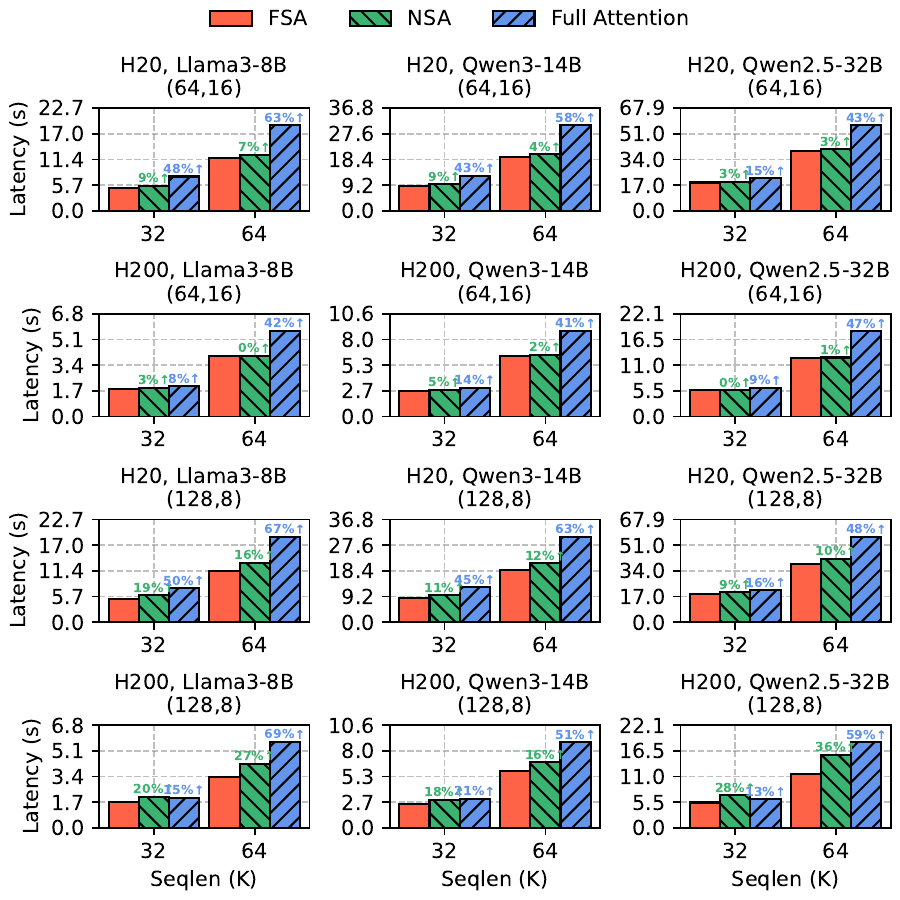}
        \caption{Inference Prefill latency of \sys, NSA, full attention.}
        \label{fig:prefill}
    \end{minipage}
\end{figure}

\subsection{End-to-end Performance Comparison (Q2)}
\label{sec:e2e}

\textbf{End-to-end training performance.} We benchmark end-to-end training performance of \sys against NSA and full attention across various models and hardware setups. As shown in Figure \ref{fig:e2e}, results demonstrate that \sys consistently reduces training latency across all evaluated cases. Specifically, \sys achieves up to 1.25$\times$ speedup and on average 1.09$\times$ speedup compared to NSA, and delivers up to 2.47$\times$ speedup and an average of 1.86$\times$ speedup compared to full attention. These efficiency gains are pronounced with longer sequences and on higher-performance hardware like the H200, demonstrating \sys's effectiveness in accelerating computation-intensive training scenarios.

\textbf{Inference performance.} For prefill latency, we benchmark \sys against NSA and full attention across various models and hardware setups. As shown in Figure \ref{fig:prefill}, our results demonstrate that \sys achieves lower prefill latency across most evaluated configurations. Specifically, \sys achieves up to 1.36$\times$ speedup and on average 1.11$\times$ speedup compared to NSA. \sys performance advantages are even more significant when compared to full attention, where \sys delivers up to 1.69$\times$ speedup and an average of 1.39$\times$ speedup. Taken together, these results underscore \sys's efficacy in accelerating the prefill phase of LLM inference.\footnote{We present more detailed decoding evaluations in Appendix \ref{app:flashdecode}.} In terms of decoding latency, \sys matches that of NSA, which reduces memory access of the decoding phase by only loading a sparse subset composed of compressed tokens, selected tokens, and recent tokens from a sliding window~\citep{yuan2025native}.

\subsection{Performance Breakdown \& Ablation Studies (Q3)}
\label{sec:case_abl_study}

In this section, we evaluate \sys at both kernel and end-to-end (training or inference) levels. At the kernel level, we analyze forward and backward performance separately, and examine each of the three attention mechanisms within NSA: Compression, selection, and sliding window on key/value tokens. We conduct ablation studies to assess the effectiveness of \sys kernel optimizations. We validate the implementation correctness of \sys by comparing training loss across \sys, NSA, and full attention in Appendix \ref{app:loss}.

\begin{figure}[t!]
    \centering
    \begin{minipage}[b]{0.48\textwidth}
        \centering
        \includegraphics[width=\linewidth]{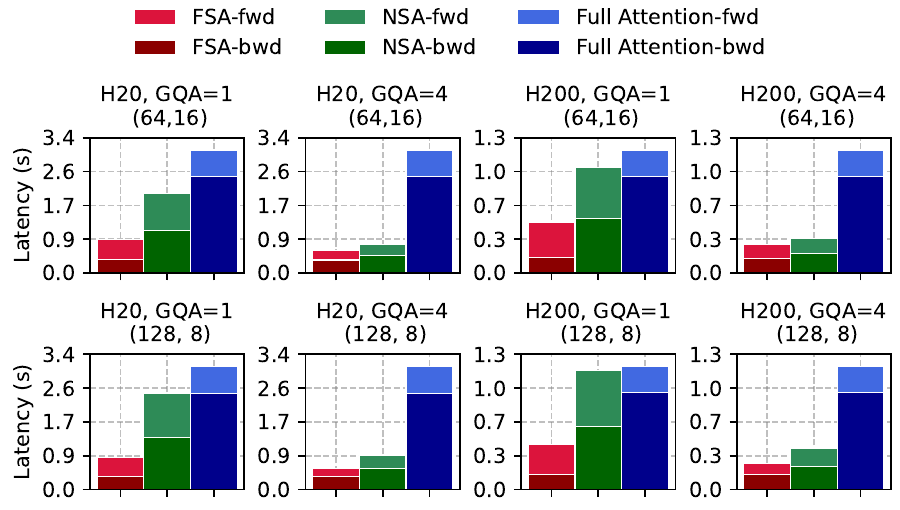}
        \caption{Experimental breakdown of \sys, NSA, and full attention latencies during forward and backward computation.}
        \label{fig:fwdbwd}
    \end{minipage}
    \hfill
    \begin{minipage}[b]{0.48\textwidth}
        \centering
        \includegraphics[width=\linewidth]{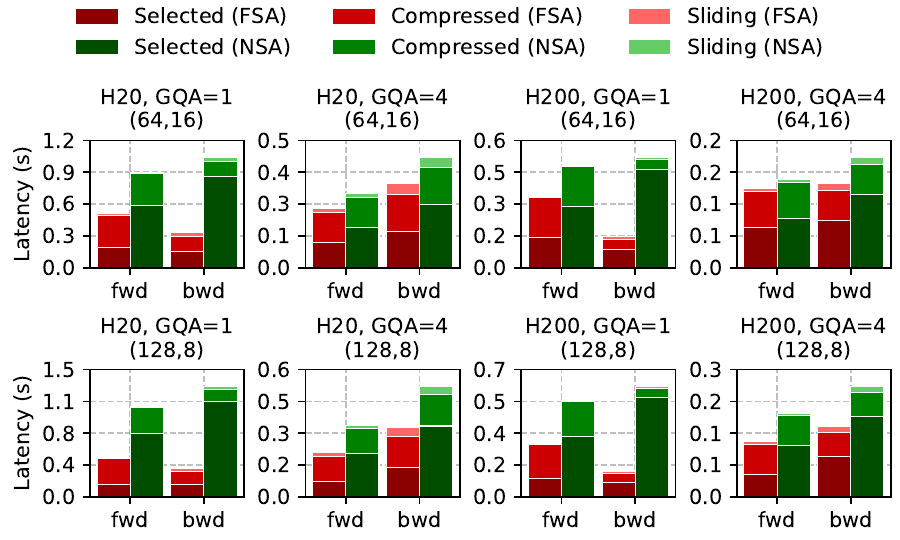}
        \caption{Experimental breakdown of token compression, selection, and sliding window attention overhead during forward/backward pass.}
        \label{fig:threephase}
    \end{minipage}
\end{figure}

\textbf{Forward and backward breakdown.} We conduct a detailed breakdown to analyze forward and backward attention computation latencies of \sys, NSA, and full attention across various NSA configurations. As shown in Figure \ref{fig:fwdbwd}, \sys demonstrates superior performance in both forward and backward attention computations across all evaluated scenarios. For forward computation, \sys achieves up to 2.36$\times$ speedup and on average 1.62$\times$ lower latency compared to NSA, and up to 3.23$\times$ speedup and on average 1.83$\times$ lower latency compared to full attention. Backward computation analysis reveals even more pronounced advantages, since \sys avoids computation costs for index tensors $\mathcal{I}_i$, $\mathcal{O}_i$ for $i$-th KV block (see details in \S\ref{sec:fsa-impl}). \sys achieves up to 4.32$\times$ speedup and on average 2.59$\times$ lower latency compared to NSA, and up to 7.45$\times$ speedup and on average 6.89$\times$ lower latency compared to full attention. Performance improvements remain consistent across different NSA configurations, demonstrating that \sys provides robust efficiency gains.

\textbf{Compression, selection, and sliding window breakdown.} We conduct detailed breakdown experiments for the three essential steps in NSA. As demonstrated in Figure \ref{fig:threephase}, the token selection phase dominates overall attention computation performance, accounting for up to 79\% and on average 65\% of total attention overhead across all evaluated configurations. And \sys achieves substantial performance improvements in token selection, delivering up to 7.6$\times$ speedup and on average 3.4$\times$ lower latency compared to NSA in this critical phase. These results highlight that \sys's primary performance advantages stem from its efficient handling of token selection computation.

\begin{figure}[t!]
    \centering
    \begin{minipage}[b]{0.48\textwidth}
        \centering
        \includegraphics[width=\linewidth]{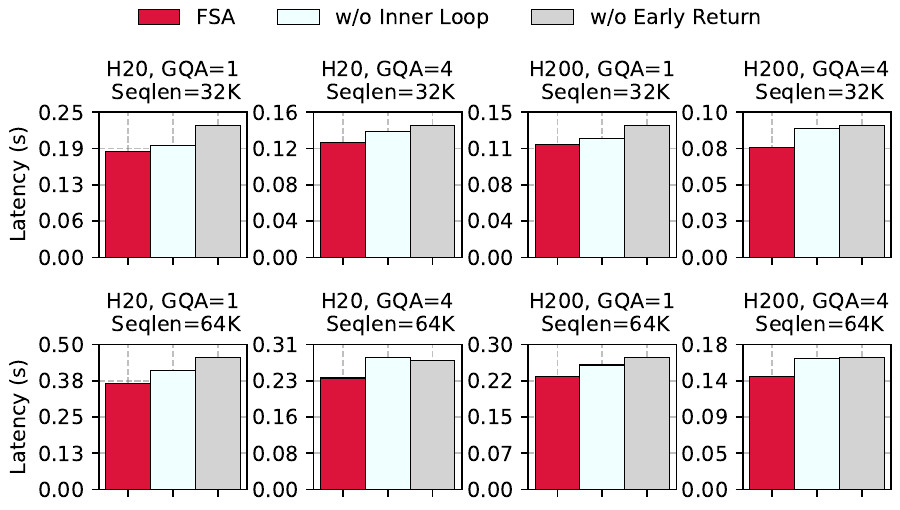}
        \caption{Ablation study (with or without \sys optimizations) on \sys kernel.}
        \label{fig:abla}
    \end{minipage}
    \hfill
    \begin{minipage}[b]{0.48\textwidth}
        \centering
        \includegraphics[width=\linewidth]{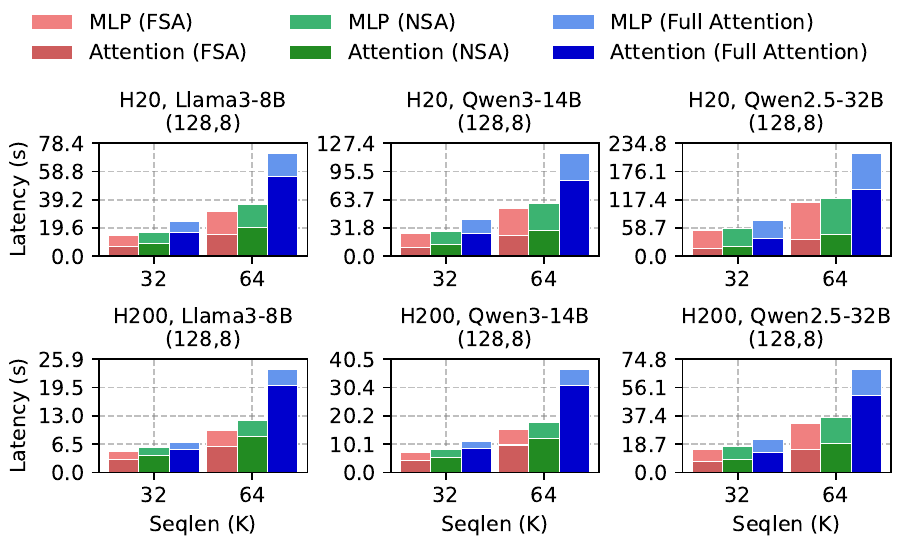}
        \caption{Breakdown of computation time for attention and MLP during end-to-end training.}
        \label{fig:e2ebk}
    \end{minipage}
\end{figure}


\textbf{Ablation study on sparse attention performance.} We present an ablation study of \sys kernel performance in Figure \ref{fig:abla}, where we disable each of additional optimizations of \sys we mentioned in~\S\ref{sec:fsa}. Results demonstrate that by disabling the inner loop (one thread block for one query batch), performance of \sys kernel drops by up to 18.9\% and on average 11.9\%, and by disabling early return optimization, performance drops by up to 25.2\% and on average 18.2\%. These empirical results demonstrate the importance of each component of our \sys optimization in enhancing performance.

\textbf{End-to-end training breakdown.} To isolate the source of performance improvements, we conduct a breakdown analysis of the end-to-end training latency. As shown in Figure \ref{fig:e2ebk}, results demonstrate that \sys's performance improvements originate from attention computation. Within this component, \sys achieves up to 1.4$\times$ and on average 1.23$\times$ lower latency than NSA, and realizes a speedup of up to 3.87$\times$ and on average 2.91$\times$ over full attention. This analysis confirms that overall end-to-end speedup is driven by \sys's fundamental optimizations in NSA token selection.


\section{Conclusion}
\label{sec:conclusion}


We presented Flash Sparse Attention (\sys), a kernel design that broadens the applicability of Native Sparse Attention (NSA) to modern LLMs where each GQA group contains a small number of query heads. By inverting kernel loop order and introducing tailored optimizations for non-contiguous memory access, online softmax, and accumulation, \sys eliminates padding inefficiencies that limit NSA on current GPUs. Evaluation demonstrates that \sys achieves substantial improvements in both kernel-level and end-to-end performance, offering consistent speedups in training/inference across state-of-the-art long-context LLMs. These results highlight that algorithm–system co-design is critical for translating theoretical efficiency of sparse attention into practical acceleration. We believe \sys provides a foundation for future exploration of hardware-efficient sparse attention.


\nocite{zhang2025efficient,zhangli2025efficient,jianghexgen,jiang2024hexgen,yan2024hexiscale,agarwal2024chai, jiang2024minference,xiao2024infllm}

\section*{Acknowledgment}

This work is supported by the HKUST startup grant R9895
from CSE; RGC-ECS project 26218024; RGC-NSFC project CRS\_HKUST601/24. 

\bibliography{iclr2026_conference}

@article{yuan2025native,
  title={Native sparse attention: Hardware-aligned and natively trainable sparse attention},
  author={Yuan, Jingyang and Gao, Huazuo and Dai, Damai and Luo, Junyu and Zhao, Liang and Zhang, Zhengyan and Xie, Zhenda and Wei, YX and Wang, Lean and Xiao, Zhiping and others},
  journal={arXiv preprint arXiv:2502.11089},
  year={2025}
}

@article{lu2025moba,
  title={Moba: Mixture of block attention for long-context llms},
  author={Lu, Enzhe and Jiang, Zhejun and Liu, Jingyuan and Du, Yulun and Jiang, Tao and Hong, Chao and Liu, Shaowei and He, Weiran and Yuan, Enming and Wang, Yuzhi and others},
  journal={arXiv preprint arXiv:2502.13189},
  year={2025}
}

@article{dao2023flashattention,
  title={Flashattention-2: Faster attention with better parallelism and work partitioning},
  author={Dao, Tri},
  journal={arXiv preprint arXiv:2307.08691},
  year={2023}
}

@article{dubey2024llama,
  title={The llama 3 herd of models},
  author={Dubey, Abhimanyu and Jauhri, Abhinav and Pandey, Abhinav and Kadian, Abhishek and Al-Dahle, Ahmad and Letman, Aiesha and Mathur, Akhil and Schelten, Alan and Yang, Amy and Fan, Angela and others},
  journal={arXiv e-prints},
  pages={arXiv--2407},
  year={2024}
}

@article{yang2025qwen3,
  title={Qwen3 technical report},
  author={Yang, An and Li, Anfeng and Yang, Baosong and Zhang, Beichen and Hui, Binyuan and Zheng, Bo and Yu, Bowen and Gao, Chang and Huang, Chengen and Lv, Chenxu and others},
  journal={arXiv preprint arXiv:2505.09388},
  year={2025}
}

@Misc{nativesparseattention,
  title = {Native Sparse Attention},
  author = {FLA Organization},
  howpublished = {\url{https://github.com/fla-org/native-sparse-attention}},
  year = {2024},
}

@Misc{nvidia_cuda_best_practices,
  title={CUDA C++ Best Practices Guide},
  author={NVIDIA},
  year={2024},
  howpublished = {\url{https://docs.nvidia.com/cuda/cuda-c-best-practices-guide/}},
}

@inproceedings{kwon2023efficient,
  title={Efficient memory management for large language model serving with pagedattention},
  author={Kwon, Woosuk and Li, Zhuohan and Zhuang, Siyuan and Sheng, Ying and Zheng, Lianmin and Yu, Cody Hao and Gonzalez, Joseph and Zhang, Hao and Stoica, Ion},
  booktitle={Proceedings of the 29th symposium on operating systems principles},
  pages={611--626},
  year={2023}
}

@article{ainslie2023gqa,
  title={Gqa: Training generalized multi-query transformer models from multi-head checkpoints},
  author={Ainslie, Joshua and Lee-Thorp, James and De Jong, Michiel and Zemlyanskiy, Yury and Lebr{\'o}n, Federico and Sanghai, Sumit},
  journal={arXiv preprint arXiv:2305.13245},
  year={2023}
}

@article{zhang2024sageattention,
  title={Sageattention: Accurate 8-bit attention for plug-and-play inference acceleration},
  author={Zhang, Jintao and Wei, Jia and Huang, Haofeng and Zhang, Pengle and Zhu, Jun and Chen, Jianfei},
  journal={arXiv preprint arXiv:2410.02367},
  year={2024}
}

@article{zhang2024sageattention2,
  title={Sageattention2: Efficient attention with thorough outlier smoothing and per-thread int4 quantization},
  author={Zhang, Jintao and Huang, Haofeng and Zhang, Pengle and Wei, Jia and Zhu, Jun and Chen, Jianfei},
  journal={arXiv preprint arXiv:2411.10958},
  year={2024}
}

@article{zhang2025sageattention3,
  title={Sageattention3: Microscaling fp4 attention for inference and an exploration of 8-bit training},
  author={Zhang, Jintao and Wei, Jia and Zhang, Pengle and Xu, Xiaoming and Huang, Haofeng and Wang, Haoxu and Jiang, Kai and Zhu, Jun and Chen, Jianfei},
  journal={arXiv preprint arXiv:2505.11594},
  year={2025}
}

@article{vaswani2017attention,
  title={Attention is all you need},
  author={Vaswani, Ashish and Shazeer, Noam and Parmar, Niki and Uszkoreit, Jakob and Jones, Llion and Gomez, Aidan N and Kaiser, {\L}ukasz and Polosukhin, Illia},
  journal={Advances in neural information processing systems},
  volume={30},
  year={2017}
}

@article{milakov2018online,
  title={Online normalizer calculation for softmax},
  author={Milakov, Maxim and Gimelshein, Natalia},
  journal={arXiv preprint arXiv:1805.02867},
  year={2018}
}

@Misc{tritonattention,
  title = {Fused Attention Tutorial},
  author = {Triton},
  year = {2024},
  howpublished = {\url{https://triton-lang.org/main/getting-started/tutorials/06-fused-attention.html}},
}

@article{team2024qwen2,
  title={Qwen2 technical report},
  author={Team, Qwen},
  journal={arXiv preprint arXiv:2407.10671},
  year={2024}
}

@Misc{h200specs,
  title = {H200 Tensor Core GPU},
  author = {NVIDIA},
  howpublished = {\url{https://www.nvidia.com/en-us/data-center/h200/}},
  year = {2024},
}

@Misc{atomic_add,
  title={CUDA C++ Programming Guide},
  author={NVIDIA},
  year={2024},
  howpublished = {\url{https://docs.nvidia.com/cuda/cuda-c-programming-guide/}},
  note={Section on Atomic Functions}
}

@misc{h20,
  title={NVIDIA H20 Solution Brief},
  author={NVIDIA},
  year={2024},
  howpublished = {\url{https://images.nvidia.com/content/pdf/dgx-apps/NVIDIA-H2O-Solution-Brief-June17.pdf}},
}

@article{shoeybi2019megatron,
  title={Megatron-lm: Training multi-billion parameter language models using model parallelism},
  author={Shoeybi, Mohammad and Patwary, Mostofa and Puri, Raul and LeGresley, Patrick and Casper, Jared and Catanzaro, Bryan},
  journal={arXiv preprint arXiv:1909.08053},
  year={2019}
}

@inproceedings{Zhao2019ExplicitSparseTransformer,
  title={Explicit Sparse Transformer: Concentrated Attention Through Explicit Selection},
  author={Zhao, Guangxiang and Lin, Junyang and Zhang, Zhiyuan and Ren, Xuancheng and Su, Qi and Sun, Xu},
  booktitle={arXiv preprint arXiv:1912.11637},
  year={2019}
}

@article{zhangli2025efficient,
  title={Efficient Mixed-Precision Large Language Model Inference with TurboMind},
  author={Zhang, Li and Jiang, Youhe and He, Guoliang and Chen, Xin and Lv, Han and Yao, Qian and Fu, Fangcheng and Chen, Kai},
  journal={arXiv preprint arXiv:2508.15601},
  year={2025}
}

@inproceedings{jiang2024hexgen,
  title={HEXGEN: generative inference of large language model over heterogeneous environment},
  author={Jiang, Youhe and Yan, Ran and Yao, Xiaozhe and Zhou, Yang and Chen, Beidi and Yuan, Binhang},
  booktitle={Proceedings of the 41st International Conference on Machine Learning},
  pages={21946--21961},
  year={2024}
}

@inproceedings{jianghexgen,
  title={HexGen-2: Disaggregated Generative Inference of LLMs in Heterogeneous Environment},
  author={JIANG, YOUHE and Yan, Ran and Yuan, Binhang},
  booktitle={The Thirteenth International Conference on Learning Representations}
}

@article{Tay2020SparseSinkhornAttention,
  title={Sparse Sinkhorn Attention},
  author={Tay, Yi and Bahri, Dara and Yang, Liu and Metzler, Donald and Juan, Da-Cheng},
  journal={arXiv preprint arXiv:2002.11296},
  year={2020}
}

@article{Lee2023SEA,
  title={SEA: Sparse Linear Attention with Estimated Attention Mask},
  author={Lee, Heejun and Kim, Jina and Willette, Jeffrey and Hwang, Sung Ju},
  journal={arXiv preprint arXiv:2310.01777},
  year={2023}
}

@Misc{ptx_warp_level_matrix,
  title  = {Parallel Thread Execution ISA Version 9.0 --- Warp-Level Matrix Instructions},
  author = {NVIDIA},
  howpublished = {\url{https://docs.nvidia.com/cuda/parallel-thread-execution/index.html#warp-level-matrix-instructions}},
  year   = {2025},
}

@inproceedings{tang2024quest,
  title={QUEST: query-aware sparsity for efficient long-context LLM inference},
  author={Tang, Jiaming and Zhao, Yilong and Zhu, Kan and Xiao, Guangxuan and Kasikci, Baris and Han, Song},
  booktitle={Proceedings of the 41st International Conference on Machine Learning},
  pages={47901--47911},
  year={2024}
}

@inproceedings{xiaoduoattention,
  title={DuoAttention: Efficient Long-Context LLM Inference with Retrieval and Streaming Heads},
  author={Xiao, Guangxuan and Tang, Jiaming and Zuo, Jingwei and Yang, Shang and Tang, Haotian and Fu, Yao and Han, Song and others},
  booktitle={The Thirteenth International Conference on Learning Representations},
  year={2024}
}

@article{zhu2024sampleattention,
  title={Sampleattention: Near-lossless acceleration of long context llm inference with adaptive structured sparse attention},
  author={Zhu, Qianchao and Duan, Jiangfei and Chen, Chang and Liu, Siran and Li, Xiuhong and Feng, Guanyu and Lv, Xin and Cao, Huanqi and Chuanfu, Xiao and Zhang, Xingcheng and others},
  journal={arXiv preprint arXiv:2406.15486},
  year={2024}
}

@article{laiflexprefill,
  title={FlexPrefill: A Context-Aware Sparse Attention Mechanism for Efficient Long-Sequence Inference},
  author={Lai, Xunhao and Lu, Jianqiao and Luo, Yao and Ma, Yiyuan and Zhou, Xun},
  booktitle={The Fourteenth International Conference on Learning Representations},
  year={2025}
}

@inproceedings{xu2025xattention,
  title     = {XAttention: Block Sparse Attention with Antidiagonal Scoring},
  author    = {Xu, Ruyi and Xiao, Guangxuan and Huang, Haofeng and Guo, Junxian and Han, Song},
  booktitle = {Proceedings of the 42nd International Conference on Machine Learning (ICML)},
  year      = {2025}
}

@article{zhang2023h2o,
  title={H2o: Heavy-hitter oracle for efficient generative inference of large language models},
  author={Zhang, Zhenyu and Sheng, Ying and Zhou, Tianyi and Chen, Tianlong and Zheng, Lianmin and Cai, Ruisi and Song, Zhao and Tian, Yuandong and R{\'e}, Christopher and Barrett, Clark and others},
  journal={Advances in Neural Information Processing Systems},
  volume={36},
  pages={34661--34710},
  year={2023}
}

@inproceedings{agarwal2024chai,
  title={CHAI: clustered head attention for efficient LLM inference},
  author={Agarwal, Saurabh and Acun, Bilge and Hosmer, Basil and Elhoushi, Mostafa and Lee, Yejin and Venkataraman, Shivaram and Papailiopoulos, Dimitris and Wu, Carole-Jean},
  booktitle={Proceedings of the 41st International Conference on Machine Learning},
  pages={291--312},
  year={2024}
}

@article{jiang2024minference,
  title={Minference 1.0: Accelerating pre-filling for long-context llms via dynamic sparse attention},
  author={Jiang, Huiqiang and Li, Yucheng and Zhang, Chengruidong and Wu, Qianhui and Luo, Xufang and Ahn, Surin and Han, Zhenhua and Abdi, Amir H and Li, Dongsheng and Lin, Chin-Yew and others},
  journal={Advances in Neural Information Processing Systems},
  volume={37},
  pages={52481--52515},
  year={2024}
}

@article{xiao2024infllm,
  title={Infllm: Training-free long-context extrapolation for llms with an efficient context memory},
  author={Xiao, Chaojun and Zhang, Pengle and Han, Xu and Xiao, Guangxuan and Lin, Yankai and Zhang, Zhengyan and Liu, Zhiyuan and Sun, Maosong},
  journal={Advances in Neural Information Processing Systems},
  volume={37},
  pages={119638--119661},
  year={2024}
}

@misc{shorten2024mlarxivpapers,
  title={ML-ArXiv-Papers},
  author={Shorten, Connor},
  year={2024},
  howpublished = {\url{https://huggingface.co/datasets/CShorten/ML-ArXiv-Papers}},
}

@article{zhang2025efficient,
  title={A Survey of Efficient Attention Methods: Hardware-efficient, Sparse, Compact, and Linear Attention},
  author={Zhang, Jintao and Su, Rundong and Liu, Chunyu and Wei, Jia and Wang, Ziteng and Zhang, Pengle and Wang, Haoxu and Jiang, Huiqiang and Huang, Haofeng and Xiang, Chendong and Xi, Haocheng and Yang, Shuo and Li, Xingyang and Hu, Yuezhou and Fu, Tianyu and Zhao, Tianchen and Zhang, Yicheng and Jiang, Youhe and Chen, Chang and Jiang, Kai and Chen, Huayu and Zhao, Min and Xu, Xiaoming and Zhu, Jun and Chen, Jianfei},
  year={2025}
}

@article{yan2024hexiscale,
  title={HexiScale: Accommodating Large Language Model Training over Heterogeneous Environment},
  author={Yan, Ran and Jiang, Youhe and Nie, Xiaonan and Fu, Fangcheng and Cui, Bin and Yuan, Binhang},
  journal={arXiv preprint arXiv:2409.01143},
  year={2024}
}

@Misc{gpt4o,
title = {OpenAI GPT-4o},
author={OpenAI},
url={https://platform.openai.com/docs/models/gpt-4o},
year = {2024},
}

@Misc{flashdecoding,
title = {Flash-Decoding for long-context inference
},
author={Tri Dao and Daniel Haziza and Francisco Massa and Grigory Sizov},
url={https://pytorch.org/blog/flash-decoding/},
year = {2023},
}

@Misc{longbench,
title = {LongBench: A Benchmark for Long-Context Language Models.},
author={Zai, Organization},
url={https://huggingface.co/datasets/zai-org/LongBench},
year = {2023},
}

@Misc{claude3,
title = {The Claude 3 Model Family: Opus, Sonnet, Haiku},
author={Anthropic},
url={https://www-cdn.anthropic.com/de8ba9b01c9ab7cbabf5c33b80b7bbc618857627/Model_Card_Claude_3.pdf},
year = {2024},
}

@article{young2024yi,
  title={Yi: Open foundation models by 01. ai},
  author={Young, Alex and Chen, Bei and Li, Chao and Huang, Chengen and Zhang, Ge and Zhang, Guanwei and Li, Heng and Zhu, Jiangcheng and Chen, Jianqun and Chang, Jing and others},
  journal={arXiv preprint arXiv:2403.04652},
  year={2024}
}

@article{xu2025128k,
  title={From 128k to 4m: Efficient training of ultra-long context large language models},
  author={Xu, Chejian and Ping, Wei and Xu, Peng and Liu, Zihan and Wang, Boxin and Shoeybi, Mohammad and Li, Bo and Catanzaro, Bryan},
  journal={arXiv preprint arXiv:2504.06214},
  year={2025}
}

@article{chen2024core,
  title={Core context aware transformers for long context language modeling},
  author={Chen, Yaofo and You, Zeng and Zhang, Shuhai and Li, Haokun and Li, Yirui and Wang, Yaowei and Tan, Mingkui},
  journal={arXiv preprint arXiv:2412.12465},
  year={2024}
}

@article{acharya2024star,
  title={Star attention: Efficient llm inference over long sequences},
  author={Acharya, Shantanu and Jia, Fei and Ginsburg, Boris},
  journal={arXiv preprint arXiv:2411.17116},
  year={2024}
}

@article{wang2024beyond,
  title={Beyond the limits: A survey of techniques to extend the context length in large language models},
  author={Wang, Xindi and Salmani, Mahsa and Omidi, Parsa and Ren, Xiangyu and Rezagholizadeh, Mehdi and Eshaghi, Armaghan},
  journal={arXiv preprint arXiv:2402.02244},
  year={2024}
}
\bibliographystyle{iclr2026_conference}

\clearpage

\appendix

\section{The Use of Large Language Models}
In this paper, we leverage LLMs to enhance academic writing quality by ensuring grammatical correctness and improving sentence structure. 

\section{Notations}
The notations used in this paper are summarized in Table \ref{tab:notations}.

\begin{table}[h!]
\centering
\caption{Notations and Explanations.}
\label{tab:notations}
\resizebox{0.8\columnwidth}{!}{%
\begin{tabular}{c | l}
\hline
\textbf{Notation} & \textbf{Explanation} \\
\hline
$N$ & Sequence length. \\
\hline
$d_K$ & Head dimension for query and key tensor. \\
\hline
$d_V$ & Head dimension for value tensor. \\
\hline
$d$ & Uniform head dimension, i.e., $d=d_K=d_V$. \\
\hline
$h$ & Number of Q heads. \\
\hline
$h_K$ & Number of KV heads. \\
\hline
$g$ & GQA group size, defined as $g =  \frac{h}{h_K}$. \\
\hline
\multirow{2}{*}{$T$} & Number of selected KV blocks of each query token. \\
& (Hyperparameter of the NSA sparse attention module.) \\
\hline
$B_K$ & Block size of each KV block; a NSA hyperparameter. \\
\hline
$b$ & Number of KV blocks; $b = \frac{N }{ B_K }$. \\
\hline

$B_Q$ & Query batch size in FSA; a FSA hyperparameter. \\
\hline
\multirow{2}{*}{$\mathcal{I}_i$} & The set of query indices attending to the $i$-th KV block. \\
& ($\mathcal{I}_i$ contain non-contiguous query indices, usually $|\mathcal{I}_i|\leq N$.) \\
\hline
\multirow{2}{*}{$\mathcal{O}_i$} & The output tensor mapping for the $i$-th KV block; e.g., $\mathcal{O}_i[j]$  \\
& gives the storage position of token $j$ in the output buffer. \\
\hline
$N_{\text{valid}}$ & The number of valid query tokens in $\mathcal{I}_i$. \\
\hline
$\mathbf{T}$ & Sparse selected KV block indices in NSA. \\
\hline
$\mathbf{Q}$,$\mathbf{KV}$ & Full query, key, and value tensor for attention computation. \\
\hline
\multirow{2}{*}{$\mathbf{Q}_{\text{batch}}$} & Non-contiguous Query batches introduced in FSA. \\
& (One thread block processes multiple $\mathbf{Q}_{\text{batch}}$.)\\
\hline
$\mathbf{K}_i$,$\mathbf{V}_{i}$ & The $i$-th KV block with $B_K$ contiguous KV tokens. \\
\hline
\multirow{2}{*}{$\mathbf{O}_{\text{buf}}$} & Intermediate buffer which holds query attention results \\
& without scaling with online softmax in FSA. \\
\hline
\end{tabular}%
}
\vspace{-1em}
\end{table}

\section{\sys Implementation Details}

\sys is implemented using 10K lines of Python and Triton code. To optimize system performance: (\underline{i}) We apply fine-grained control over \sys selected attention kernel and reduction kernel to optimize warp-level parallelism. \sys usually assigns 4 warps per thread block for \sys selected attention kernel, which contains matrix multiplication operations, to enable sufficient computational resources of a given thread block. \sys usually assigns 1 to 2 warps per thread block for reduction kernel, which mainly consists of elementwise operations. Warp assignment for reduction kernel efficiently utilizes warp-level parallelism, reducing reduction kernel execution latency. (\underline{ii}) We speculatively compute online softmax statistics once per KV heads. Due to invariant nature of online softmax~\citep{milakov2018online}, correctness of \sys is maintained, while significant cost for computing online softmax statistics is amortized.

\section{\sys Correctness}
\label{app:loss}

\begin{figure}[t!]
    \centering
    \includegraphics[width=0.9\linewidth]{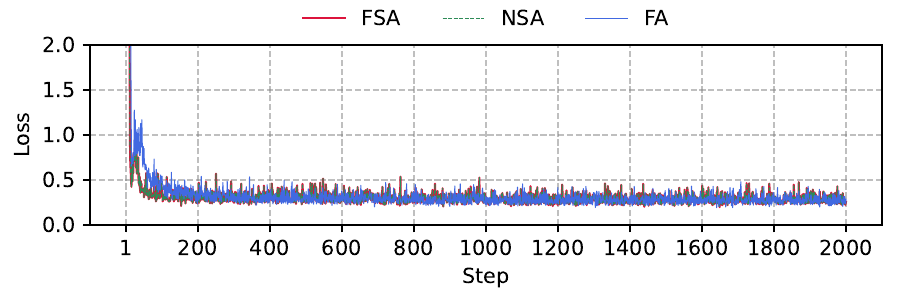}
    \vspace{-0.5em}
    \caption{Loss comparison of \sys/NSA/full attention in end-to-end Llama3-8B training.}
    \label{fig:loss}
\end{figure}

\textbf{\sys correctness.} To evaluate correctness of \sys kernels, we fine-tune Llama3-8B model using ML-ArXiv-Papers dataset~\citep{shorten2024mlarxivpapers}. We replace attention module of Llama3-8B model with either \sys or NSA, while initializing all other components with pretrained model checkpoints provided by Meta. For fair comparison with full attention, we reinitialize the parameters of the attention module. Loss comparison among \sys, NSA, and full attention is presented in Figure \ref{fig:loss}. Results demonstrate that all three methods achieve stable and similar convergence, and \sys exhibits a similar loss curve to NSA, validating the correctness of the \sys kernel.

\ryan{
\textbf{Further analysis.} To strengthen our accuracy evaluation, we conduct additional experiments by fine-tuning smaller models across diverse tasks. Specifically, we fine-tune Llama-3.2-1B, Llama-3.2-3B, and Llama-3.1-8B~\cite{dubey2024llama} on three representative LongBench~\cite{longbench} tasks:  multi-document QA (MQA) on the HotpotQA dataset, single-document QA (SQA) on the Qasper dataset, and synthetic-data QA (Synthetic) on the PassR-EN dataset. For Llama-3.2-1B and Llama-3.2-3B, we report the average loss after convergence over 2K training steps, comparing FSA, NSA, and Full Attention; these results appear in Table \ref{tab:losses}. To more comprehensively assess accuracy preservation, we further evaluate perplexity and QA F1 across all three models and tasks. The results, summarized in Tables \ref{tab:ppl} and \ref{tab:qaf1}, consistently show that FSA matches the accuracy of NSA and Full Attention.

\begin{table}[ht!]
\caption{\ryan{Converged Loss Across Datasets and Attention Modes.}}
\label{tab:losses}
\centering
\ryan{
\begin{tabular}{c|ccc|ccc|ccc}
\hline
\multirow{2}{*}{Model Size} & \multicolumn{3}{c|}{MQA (HQA)} & \multicolumn{3}{c|}{SQA (Qasper)} & \multicolumn{3}{c}{Synthetic (PassR-EN)} \\
 & FA & FSA & NSA & FA & FSA & NSA & FA & FSA & NSA \\
\hline
\multirow{1}{*}{1B} & 0.200 & 0.182 & 0.187 & 0.216 & 0.191 & 0.184 & 0.231 & 0.224 & 0.231 \\

\multirow{1}{*}{3B} & 0.173 & 0.153 & 0.166 & 0.087 & 0.082 & 0.078 & 0.123 & 0.119 & 0.118 \\

\hline
\end{tabular}
}
\end{table}

\begin{table}[ht!]
\caption{\ryan{PPL Across Datasets, Models, and Attention Modes.}}
\label{tab:ppl}
\centering
\ryan{
\begin{tabular}{c|ccc|ccc|ccc}
\hline
\multirow{2}{*}{Model Size} 
& \multicolumn{3}{c|}{MQA (HQA)} 
& \multicolumn{3}{c|}{SQA (Qasper)} 
& \multicolumn{3}{c}{Synthetic (PassR-EN)} \\
& FA & FSA & NSA & FA & FSA & NSA & FA & FSA & NSA \\
\hline
1B & 5.40 & 6.79 & 6.82 & 8.77 & 9.48 & 9.45 & 3.48 & 2.52 & 2.49 \\
3B & 2.42 & 1.50 & 1.48 & 1.20 & 2.64 & 2.62 & 1.87 & 1.94 & 1.96 \\
8B & 1.57 & 1.17 & 1.16 & 1.28 & 1.71 & 1.70 & 1.21 & 1.26 & 1.27 \\
\hline
\end{tabular}
}
\end{table}

\begin{table}[ht!]
\caption{\ryan{QA F1 Across Datasets, Models, and Attention Modes.}}
\label{tab:qaf1}
\centering
\ryan{
\begin{tabular}{c|ccc|ccc|ccc}
\hline
\multirow{2}{*}{Model Size} 
& \multicolumn{3}{c|}{MQA (HQA)} 
& \multicolumn{3}{c|}{SQA (Qasper)} 
& \multicolumn{3}{c}{Synthetic (PassR-EN)} \\
& FA & FSA & NSA & FA & FSA & NSA & FA & FSA & NSA \\
\hline
1B 
& 0.05 & 0.10 & 0.11 
& 0.08 & 0.07 & 0.06 
& 0.22 & 0.32 & 0.31 \\
3B 
& 0.28 & 0.35 & 0.33 
& 0.15 & 0.11 & 0.12 
& 0.39 & 0.47 & 0.48 \\
8B
& 0.32 & 0.38 & 0.37
& 0.23 & 0.20 & 0.19
& 0.83 & 0.86 & 0.86 \\
\hline
\end{tabular}
}
\end{table}
}

\section{\sys and NSA Theoretical Memory Access and FLOPs Analysis}
\label{app:memory}

To demonstrate how \sys outperforms NSA selected attention, we analyze as follows. For simplicity, we assume query/key/value have the same head dimension, i.e.  $d=d_K=d_V$.

\textbf{\sys analytic advantages.} \textit{Theoretically, \sys introduces lower memory access volume and number of floating-point operations (FLOPs) for small GQA group sizes}. We analyze \sys/NSA as follows:

\textit{\sys memory access volume and FLOPs.} We analyze the three key components in \sys as follows: 

\begin{itemize}[topsep=5pt, leftmargin=*]
\vspace{-0.5em}

\item \textbf{\sys selected attention kernel} launches $hb$ thread blocks, where $h$ is the number of query attention heads, and $b$ is the total number of KV blocks. For a sequence of $N$ tokens, the number of KV blocks $b=\frac{N}{B_K}$, where $B_K$ is the KV block size. In one thread block, \sys selected attention kernel runs a two-level loop. In the outer loop, it loads $2 B_K d$ KV tokens; in the inner loop, it iteratively loads $B_Qd$ query tokens, performs attention computation with a FLOPs of $4B_Q B_K d$, and stores $B_Q d$ query attention results. We estimate the number of our inner loop as follows. Assume each query token attends to each KV block with equal probability. Therefore, each query token attends to a given KV block with a probability of $\frac{T}{b}$, resulting in an average number of tokens attending to a given KV block of $\frac{N T}{b}$, and an average number of query batches for one KV block of $\frac{N T}{b B_Q}$. Assuming each data occupies $2$ bytes, we can calculate memory accessed in bytes by \sys selected attention kernel as $4dhN  (1 + T)$, and FLOPs as $4 dhN B_K T$.

\item \textbf{\sys online softmax kernel} operates similarly to the \sys selected attention kernel, with three key differences: It is called per KV head, omits V tensor loading and computation, and intermediate attention scores storage, storing only a single scalar value per (query token, KV block) pair. Following a similar estimation logic as \sys selected attention kernel, the online softmax kernel introduces $2 d h_K N (1 + T)$ memory access volume in bytes, and $2 d h_K N B_K T$ FLOPs.

\item \textbf{\sys reduction kernel} introduces negligible FLOPs, but for each query token, it involves loading attention results of $T$ KV blocks and storing the final attention results. Therefore, \sys reduction kernel introduces $2dh N  (1 + T)$ memory access in bytes. 

In total, \sys incurs $dN (6h + 2h_K) (1 + T)$ memory access in bytes, and $dN B_K T (4h+2h_K) $ FLOPs.
\vspace{-0.5em}
\end{itemize}
\vspace{-0.5em}

\textit{NSA Memory access volume and FLOPs.} NSA selected attention kernel launches $h_K N$ thread blocks, where $h_K$ is the number of KV heads. In each thread block, NSA kernel runs a two-level loop. In the outer loop, NSA kernel loads one query token and $g=\frac{h}{h_K}$ Q heads that share the same KV head. Due to the hardware requirements on matrix multiplication shapes, when GQA $<$ 8, NSA kernels must load 8 query heads ($8d$ elements), perform computation, and mask out the undesired computation results. In the inner loop, NSA kernel iteratively (up to $T$ times) loads one KV block ($2B_K d$ elements) and performs attention computation with a FLOPs of $32 B_K d$. To maintain the causal property, i.e., avoiding query tokens to attend to future KV tokens, the actual number of KV blocks that need to be loaded and participate in computations within a thread block is on average $\frac{T}{2}$. Finally, NSA kernel stores the attention results in the output tensor, incurring $g d$ memory access. Therefore, we can estimate the memory access volume (2 bytes per data) for NSA kernel as $2d h_K N (B_K T+ g + 8)$. The FLOPs for NSA kernel are $32  d h_K N B_K  T$.


\textit{\sys selected attention kernels exhibit lower memory access volume and FLOPs.} With $(B_K,T)=(64,16)$ and sequence length of 64K, which is the same configuration as presented in the NSA paper, we observe that compared to the NSA selected attention kernel, our method incurs lower memory access volume and FLOPs for GQA$\leq$8, detailed comparisons are presented in Figure \ref{fig:ablation}. In particular, for
GQA$=$4, a common configuration in LLMs, our method theoretically reduces memory access volume to 21.3\% and FLOPs to 56.2\% of those in NSA. Benefits from the more efficient hardware-aligned kernel design, our method substantially outperforms NSA across various GQA group sizes. Additionally, our method demonstrates superior performance as the NSA hyperparameter $B_K$ increases. This advantage stems from NSA's inherent inefficiency with larger KV blocks. Although NSA can easily skip loading KV blocks that fully violate causal property, to maintain causality constraints for KV blocks that partially violate causal property, NSA must mask out many KV tokens within the KV block, leading to wasteful memory accesses where loaded data is only partially valid for computation. As the KV block size $B_K$ grows larger, this inefficiency becomes increasingly pronounced, as a greater proportion of the loaded KV block remains unused due to causal masking. In contrast, our method processes all query tokens that attend to a given KV block within a single thread block, naturally satisfying causal constraints without requiring extensive masking. This approach achieves superior memory efficiency by ensuring that all loaded KV data contributes to the computation, resulting in significantly lower memory access overhead.


\textbf{\sys trade-offs.} \textit{\sys trades lowered memory access volume and FLOPs with non-contiguous loading and more buffer overhead.} Theoretical advantages of \sys come at the price of involving non-contiguous memory access and more buffers that occupy HBM memory. We analyze how these factors compromise \sys performance and how \sys optimizes memory access and buffer management as follows:

\begin{itemize}[topsep=5pt, leftmargin=*]
\vspace{-0.5em}

\item \textbf{Optimize memory access.} The non-contiguous loading on query batches, which is inefficient on modern GPUs, compromises \sys selected attention kernel performance.  Modern GPUs usually operate more efficiently under coalesced and contiguous memory access, which can improve the L2-cache hit rate and thereby kernel efficiency~\citep{nvidia_cuda_best_practices}. Therefore, the theoretical advantages of our method cannot be fully reflected in actual hardware, due to inevitably degraded performance of non-contiguous memory access. Nonetheless, to our best effort, \sys optimizes memory access with fine-grained early return mechanisms that filter out unnecessary query batches loading. For example, for $i$-th KV block, \sys compactly stores query indices in set $\mathcal{I}_i$, which is computed via a full index table. For each query token, the full index table records whether it should attend to $i$-th KV block, and $\mathcal{I}_i$ filters the tokens that do not attend to $i$-th KV block. Therefore, when all query tokens in $\mathcal{I}_i$ are exhausted, \sys returns early.

\item \textbf{Optimize buffer management.} The newly introduced buffers, $\mathbf{O}_{\text{buf}}$ appeared in Figure \ref{fig:kernel_design} (right), bring memory overhead. \sys minimizes buffer overhead from two aspects: (\underline{i}) \sys Token selection kernel processes a subset of query heads at each time, reusing the buffers for subsequent query heads computations. (\underline{ii}) \sys introduces an output index mapping tensor to store results compactly. For each query head, \sys only reserves buffers for maximum query tokens that attend to a given KV block. On average, this value is $B_K T$, combining that $b=\frac{N}{B_K}$, \sys introduces an output buffer with $d N T$ elements. Assume each data in the output buffer occupies $2$ bytes, for a sequence with 64K tokens, $T$ at 16, and $d$ at 128, $\mathbf{O}_{\text{buf}}$ occupies $1$ GB HBM memory (This also applies for the buffer for intermediate gradients with respect to $\mathbf{Q}$). Compared to the high HBM memory capacity in modern GPUs,e.g., $96$ GB HBM memory on H20~\citep{h20} and $141$ GB memory on H200~\citep{h200specs}, the additional buffer overhead in \sys remains manageable.
\vspace{-0.5em}

\end{itemize}
\vspace{-0.5em}

\textbf{Attention Sink Optimizations.} The attention sink phenomenon in NSA sparse token selection presents a challenge for \sys's buffer management strategy. The initial KV block receives attention from all query tokens, while subsequent KV blocks exhibit more selective attention patterns. This asymmetry creates a buffer allocation dilemma: In practice, \sys allocates uniform buffer sizes based on the maximum number of valid tokens across all KV blocks. However, the attention sink property forces this maximum as full sequence length, thereby negating the memory efficiency gains that \sys's sparse buffer management is designed to achieve. To address this inefficiency, we implement a dual-buffer allocation strategy. We maintain separate buffer allocations for the attention sink (first KV block) and the remaining KV blocks. The attention sink buffer accommodates the full query sequence, while buffers for subsequent KV blocks are sized according to their maximum valid query tokens, which are usually much smaller than full sequence length. This approach preserves the memory optimization benefits for the majority of KV blocks while handling the attention sink's dense connectivity requirements. 

\textbf{\sys online profiling module.} \textit{In real-world deployment, \sys dynamically selects kernel configuration via online profiling, and potentially falls back to original NSA implementation.} To ensure optimal performance across diverse NSA configurations, \sys incorporates a one-time online profiling mechanism. Upon its first execution with a new set of hyperparameters (e.g., sequence length, GQA group size), \sys benchmarks its kernel performance across several candidate query batch sizes (e.g., 1, 64, 128). When GQA group size is sufficiently large, a query batch size of 1 is additionally searched and serves as a potential fallback to original NSA strategy of batching query heads. Once profiling is complete, the fastest configuration is cached. All subsequent calls with the same hyperparameters directly use this optimal configuration, bypassing profiling step until hyperparameters change.

\ryan{
\textbf{Actual memory footprint of FSA buffers.} We conduct additional micro-benchmarks to measure the memory footprint of FSA buffers. Concretely, we set the head dimension at 128 and use the NSA hyperparameters ($B_K$, $T$) = (64,16) or (128,8), and report the profiled buffer overheads for sequence lengths ranging from 32K to 256K in Table \ref{tab:buffer}.  Under extreme cases, i.e., when the sequence length is 128K or 256K, FSA introduces 5.01GB or 12.36 GB buffer memory overhead, which is still much smaller than the memory capacity of modern GPUs (e.g., H200 has 141GB memory). These results confirm that the FSA buffer memory overhead remains acceptable.
}

\begin{table}[h]
\caption{\ryan{Profiled Buffer Overhead.}}
\label{tab:buffer}
\centering
\ryan{
\begin{tabular}{c c c}
\hline
($B_K$, $T$) & Seqlen (K) & Profiled Buffer Overhead (GB) \\
\hline
(64, 16)  & 32  & 0.52 \\
(64, 16)  & 64  & 1.88 \\
(64, 16)  & 128 & 5.01 \\
(64, 16)  & 256 & 12.36 \\
(128, 8)  & 32  & 0.26 \\
(128, 8)  & 64  & 0.91 \\
(128, 8)  & 128 & 2.28 \\
(128, 8)  & 256 & 6.15 \\
\hline
\end{tabular}
}
\end{table}

\ryan{
\section{Evaluations for Ultra Long Sequence Lengths.}
\label{app:ultra-long}

We extend our evaluations to 128K and 256K sequence lengths. Fixing the head dimension at 128 and the number of query heads at 64, while varying the number of key and value heads, we evaluate configurations where a GQA group contains 1 to 8 query heads. Using the NSA hyperparameters with ($B_K$, $T$) = (64, 16) or (128, 8), we benchmark the performance of FSA, NSA, and Full Attention (FA) on both H20 and H200 GPUs.

\subsection{Inference Prefill and Training Evaluations}

\textbf{Results discussion:} The experimental results in Table \ref{tab:H20-long} and \ref{tab:H200-long} show that FSA also outperforms NSA for ultra-long sequence lengths. For inference prefill execution latency, FSA achieves up to 1.47$\times$ speedup and an average of 1.20$\times$ lower kernel latency on H20 GPUs, and up to 1.86$\times$ speedup with an average of 1.23$\times$ lower kernel latency on H200 GPUs, compared to NSA. For training execution latency — measured over one forward and one backward pass — FSA achieves up to 1.91$\times$ speedup and an average of 1.37$\times$ lower kernel latency on H20 GPUs, and up to 2.55$\times$ speedup with an average of 1.49$\times$ lower kernel latency on H200 GPUs, relative to NSA.

\begin{table}[ht!]
\caption{\ryan{H20 GPU, Inference Prefill and Training Latency for Different ($B_K$, $T$).}}
\label{tab:H20-long}
\centering
\ryan{
\begin{tabular}{c c c c c c c c c}
\hline
\multirow{2}{*}{($B_K$, $T$)} & \multirow{2}{*}{GQA} & Seq Len & FSA & NSA  & FA   & FSA  & NSA & FA \\
 & & (K) & Fwd (s) & Fwd (s) & Fwd (s) & F + B (s) & F + B (s) & F + B (s) \\
\hline
\multirow{8}{*}{(64,16)}
 & 1 & 128 & 1.42 & 2.08 & 2.64 & 2.36 & 4.51 & 12.08 \\
 & 1 & 256 & 6.40 & 7.18 & 10.5 & 8.70 & 13.29 & 48.23 \\
 & 2 & 128 & 0.87 & 1.17 & 2.62 & 1.79 & 2.71 & 12.04 \\
 & 2 & 256 & 3.74 & 4.07 & 10.53 & 6.03 & 8.43 & 48.27 \\
 & 4 & 128 & 0.52 & 0.61 & 2.65 & 1.43 & 1.75 & 12.07 \\
 & 4 & 256 & 2.41 & 2.44 & 10.52 & 4.66 & 5.98 & 48.24 \\
 & 8 & 128 & 0.45 & 0.45 & 2.61 & 1.38 & 1.39 & 12.05 \\
 & 8 & 256 & 1.63 & 1.64 & 10.51 & 3.99 & 4.75 & 48.26 \\
\hline
\hline
\multirow{8}{*}{(128,8)}
 & 1 & 128 & 1.24 & 1.68 & 2.64 & 2.15 & 3.50 & 12.08 \\
 & 1 & 256 & 5.34 & 7.50 & 10.5 & 7.56 & 13.65 & 48.23 \\
 & 2 & 128 & 0.90 & 1.29 & 2.62 & 1.66 & 2.22 & 12.04 \\
 & 2 & 256 & 3.18 & 4.24 & 10.53 & 5.40 & 8.33 & 48.27 \\
 & 4 & 128 & 0.45 & 0.49 & 2.65 & 1.35 & 1.52 & 12.07 \\
 & 4 & 256 & 2.10 & 2.53 & 10.52 & 4.29 & 5.55 & 48.24 \\
 & 8 & 128 & 0.42 & 0.43 & 2.61 & 1.31 & 1.41 & 12.05 \\
 & 8 & 256 & 1.56 & 1.68 & 10.51 & 3.74 & 4.17 & 48.28 \\
\hline
\end{tabular}
}
\end{table}

\begin{table}[ht!]
\caption{\ryan{H200 GPU, Inference Prefill and Training Latency for Different ($B_K$, $T$).}}
\label{tab:H200-long}
\centering
\ryan{
\begin{tabular}{c c c c c c c c c}
\hline
\multirow{2}{*}{($B_K$, $T$)} & \multirow{2}{*}{GQA} & Seq Len & FSA & NSA  & FA   & FSA  & NSA & FA \\
 & & (K) & Fwd (s) & Fwd (s) & Fwd (s) & F + B (s) & F + B (s) & F + B (s) \\
\hline
\multirow{8}{*}{(64,16)}
 & 1 & 128 & 0.78 & 1.01 & 1.01 & 1.12 & 2.17 & 4.70 \\
 & 1 & 256 & 3.92 & 3.97 & 3.96 & 4.81 & 6.99 & 18.75 \\
 & 2 & 128 & 0.46 & 0.57 & 0.98 & 0.80 & 1.24 & 4.68 \\
 & 2 & 256 & 2.14 & 2.17 & 3.98 & 3.04 & 4.12 & 18.77 \\
 & 4 & 128 & 0.27 & 0.33 & 0.99 & 0.60 & 0.82 & 4.71 \\
 & 4 & 256 & 1.20 & 1.29 & 3.99 & 2.15 & 2.70 & 18.76 \\
 & 8 & 128 & 0.18 & 0.18 & 0.97 & 0.50 & 0.52 & 4.67 \\
 & 8 & 256 & 0.73 & 0.73 & 3.97 & 1.72 & 2.01 & 18.78 \\
\hline
\hline
\multirow{8}{*}{(128,8)}
 & 1 & 128 & 0.65 & 1.20 & 1.00 & 0.97 & 2.47 & 4.70 \\
 & 1 & 256 & 3.16 & 4.10 & 3.96 & 3.99 & 6.79 & 18.75 \\
 & 2 & 128 & 0.38 & 0.66 & 0.98 & 0.70 & 1.40 & 4.68 \\
 & 2 & 256 & 1.76 & 2.21 & 3.99 & 2.58 & 3.90 & 18.77 \\
 & 4 & 128 & 0.21 & 0.28 & 0.99 & 0.53 & 0.73 & 4.71 \\
 & 4 & 256 & 1.06 & 1.24 & 3.97 & 1.87 & 2.44 & 18.76 \\
 & 8 & 128 & 0.17 & 0.20 & 0.97 & 0.48 & 0.58 & 4.67 \\
 & 8 & 256 & 0.71 & 0.75 & 3.95 & 1.52 & 1.70 & 18.78 \\
\hline
\end{tabular}
}
\end{table}

\subsection{Inference End-to-End Evaluations}

By further fixing the number of generated tokens at 512, we evaluate the end-to-end inference execution latency of FSA, NSA, and Full Attention on both H20 and H200 GPUs.

\textbf{Results and discussion:} The experimental results in Table \ref{tab:h20-inf} and \ref{tab:h200-inf} demonstrate that FSA’s performance scales well for extremely long sequences. For inference execution latency: (i) compared to NSA, FSA achieves up to 1.40$\times$ speedup and on average 1.16$\times$ lower kernel latency on H20 GPUs, and up to 1.59$\times$ speedup and on average 1.15$\times$ lower kernel latency on H200 GPUs. (ii) Compared to Full Attention, FSA achieves up to 7.20$\times$ speedup and on average 4.61$\times$ lower kernel latency on H20 GPUs, and up to 4.71$\times$ speedup and on average 2.96$\times$ lower kernel latency on H200 GPUs.

\begin{table}[ht!]
\centering
\caption{\ryan{H20 Inference Latency (s) for ($B_K$, $T$) at (64,16) and (128,8).}}
\label{tab:h20-inf}
\ryan{
\begin{tabular}{c|c|cc|cc|cc|cc}
\hline
\multirow{2}{*}{Method} & \multirow{2}{*}{($B_K$, $T$)} 
& \multicolumn{2}{c|}{GQA = 1} & \multicolumn{2}{c|}{GQA = 2}
& \multicolumn{2}{c|}{GQA = 4} & \multicolumn{2}{c}{GQA = 8} \\
\cline{3-10}
& & 128K & 256K & 128K & 256K & 128K & 256K & 128K & 256K \\
\hline
\multirow{2}{*}{FSA} 
& (64,16)  & 1.67 & 6.71 & 1.12 & 4.05 & 0.77 & 2.72 & 0.70 & 1.94 \\
& (128,8) & 1.49 & 5.65 & 1.15 & 3.49 & 0.70 & 2.41 & 0.67 & 1.87 \\
\hline
\multirow{2}{*}{NSA} 
& (64,16)  & 2.33 & 7.49 & 1.42 & 4.38 & 0.86 & 2.75 & 0.70 & 1.95 \\
& (128,8) & 1.93 & 7.81 & 1.54 & 4.55 & 0.74 & 2.84 & 0.68 & 1.99 \\
\hline
FA & -- 
& 4.34 & 13.45 & 4.32 & 13.48 & 4.35 & 13.47 & 4.31 & 13.46 \\
\hline
\end{tabular}
}
\end{table}

\begin{table}[ht!]
\centering
\caption{\ryan{H200 Inference Latency (s) for ($B_K$, $T$) at (64,16) and (128,8).}}
\label{tab:h200-inf}
\ryan{
\begin{tabular}{c|c|cc|cc|cc|cc}
\hline
\multirow{2}{*}{Method} & \multirow{2}{*}{($B_K$, $T$)} 
& \multicolumn{2}{c|}{GQA = 1} & \multicolumn{2}{c|}{GQA = 2}
& \multicolumn{2}{c|}{GQA = 4} & \multicolumn{2}{c}{GQA = 8} \\
\cline{3-10}
& & 128K & 256K & 128K & 256K & 128K & 256K & 128K & 256K \\
\hline
\multirow{2}{*}{FSA} 
& (64,16)  & 1.06 & 4.24 & 0.74 & 2.46 & 0.55 & 1.52 & 0.46 & 1.05 \\
& (128,8) & 0.93 & 3.48 & 0.66 & 2.08 & 0.49 & 1.38 & 0.45 & 1.03 \\
\hline
\multirow{2}{*}{NSA} 
& (64,16)  & 1.29 & 4.29 & 0.85 & 2.49 & 0.61 & 1.61 & 0.46 & 1.05 \\
& (128,8) & 1.48 & 4.42 & 0.94 & 2.53 & 0.56 & 1.56 & 0.48 & 1.07 \\
\hline
FA & -- 
& 1.88 & 4.86 & 1.85 & 4.88 & 1.86 & 4.89 & 1.84 & 4.87 \\
\hline
\end{tabular}
}
\end{table}

\section{Comparison with FlashDecoding}
\label{app:flashdecode}

To compare with the state-of-the-art FlashDecoding kernel~\cite{flashdecoding}, we conducted additional experiments measuring decoding execution latency for FlashDecoding and FSA. Given a prefill sequence length (ranging from 32K to 256K), we present the average decoding latency across 1K generated tokens. We fixed the number of attention heads at 64 and the head dimension at 128. For FSA, the sparse-attention hyperparameters were set to a block size $B_K$ of 64 and TopK Value $T$ of 16. 

\textbf{Result discussions:} Experimental results in Table \ref{tab:flashdecode} demonstrate that FSA achieves superior performance to FlashDecoding. Compared to FlashDecoding, FSA demonstrates an average speedup of 5.46x on H20 GPU and 2.16x on H200 GPU. During the decoding phase, FlashDecoding partitions the key and value tokens and distributes the resulting attention computation tasks across multiple thread blocks, thereby increasing kernel-level parallelism and improving decoding throughput. However, due to the sparsity in FSA, the FSA decoding throughput is still superior to FlashDecoding.

\begin{table}[h]
\caption{\ryan{Decoding Latency on H20 and H200 GPUs.}}
\label{tab:flashdecode}
\centering
\ryan{
\begin{tabular}{c|ccc|ccc}
\hline
\multirow{2}{*}{Seq Len (K)} &
\multicolumn{3}{c|}{H20 Latency (ms)} &
\multicolumn{3}{c}{H200 Latency (ms)} \\
 & FlashDecoding & NSA & FSA & FlashDecoding & NSA & FSA \\
\hline
32  & 0.88 & 0.46 & 0.45 & 0.51 & 0.48 & 0.47 \\
64  & 1.71 & 0.46 & 0.48 & 0.88 & 0.53 & 0.54 \\
128 & 3.32 & 0.50 & 0.48 & 1.70 & 0.55 & 0.54 \\
256 & 5.76 & 0.62 & 0.61 & 1.75 & 0.62 & 0.63 \\
\hline
\end{tabular}
}
\end{table}

\section{Compliation Overhead}

To determine the optimal Triton kernel hyperparameters, both FSA and NSA incur a compilation overhead. For a given NSA hyperparameter combination, this overhead occurs only once. Setting ($B_K$, $T$) at (64, 16) or (128, 8), we evaluate the compilation overhead of FSA and NSA for sequence length across diverse sequence lengths. The experimental results are summarized in Table \ref{tab:compile}.

\begin{table}[h]
\caption{\ryan{Compilation Overhead on H20 and H200 GPUs.}}
\label{tab:compile}
\centering
\ryan{
\begin{tabular}{c c c c}
\hline
Seqlen (K) & Framework & H20 Overhead (s) & H200 Overhead (s) \\
\hline
32  & FSA & 2.16 & 1.82 \\
32  & NSA & 2.12 & 1.78 \\
64  & FSA & 2.37 & 2.01 \\
64  & NSA & 2.33 & 1.95 \\
128 & FSA & 2.59 & 2.24 \\
128 & NSA & 2.55 & 2.19 \\
256 & FSA & 2.80 & 2.36 \\
256 & NSA & 2.76 & 2.30 \\
\hline
\end{tabular}
}
\end{table}

\section{Evaluations on Distributed Performance.}
\label{app:dist}

\subsection{Distributed Inference Evaluation of the Attention Module}

\begin{table}[h]
\caption{\ryan{Distributed inference latency of the attention module on H20 GPU.}}
\label{tab:h20-dist-attn}
\centering
\ryan{
\begin{tabular}{c c c c c c c}
\hline
($B_K$, $T$) & Seq Len (K) & Framework & TP=1 (ms) & TP=2 (ms) & TP=4 (ms) & TP=8 (ms) \\
\hline
(64, 16)  & 32 & FSA & 82.50 & 45.00 & 25.94 & 16.25 \\
(64, 16)  & 32 & NSA & 99.53 & 53.44 & 28.75 & 16.56 \\
(64, 16)  & 64 & FSA & 195.84 & 110.63 & 61.25 & 38.63 \\
(64, 16)  & 64 & NSA & 221.49 & 122.81 & 65.31 & 39.94 \\
(128, 8)  & 32 & FSA & 80.31 & 43.44 & 24.69 & 15.63 \\
(128, 8)  & 32 & NSA & 105.10 & 54.38 & 28.68 & 16.75 \\
(128, 8)  & 64 & FSA & 187.50 & 102.68 & 56.25 & 33.75 \\
(128, 8)  & 64 & NSA & 243.88 & 130.31 & 70.69 & 40.00 \\
\hline
\end{tabular}
}
\end{table}

\begin{table}[h]
\caption{\ryan{Distributed inference latency of the attention module on H200 GPU.}}
\label{tab:h200-dist-attn}
\centering
\ryan{
\begin{tabular}{c c c c c c c}
\hline
($B_K$, $T$) & Seq Len (K) & Framework & TP=1 (ms) & TP=2 (ms) & TP=4 (ms) & TP=8 (ms) \\
\hline
(64, 16) & 32 & FSA & 43.44 & 25.00 & 15.63 & 11.88 \\
(64, 16) & 32 & NSA & 50.31 & 27.19 & 17.63 & 12.69 \\
(64, 16) & 64 & FSA & 110.00 & 63.13 & 39.38 & 26.13 \\
(64, 16) & 64 & NSA & 121.56 & 66.81 & 40.75 & 27.00 \\
(128, 8) & 32 & FSA & 40.63 & 23.13 & 14.38 & 10.00 \\
(128, 8) & 32 & NSA & 59.06 & 31.25 & 17.50 & 11.63 \\
(128, 8) & 64 & FSA & 96.25 & 53.75 & 32.50 & 22.81 \\
(128, 8) & 64 & NSA & 124.38 & 65.69 & 37.50 & 25.56 \\
\hline
\end{tabular}
}
\end{table}

We conduct additional experiments to evaluate the distributed inference performance of the attention module using FSA and NSA on H20 and H200 GPUs. We fix the number of query heads at 32, and the number of key and value heads at 8. This setting indicates that one GQA group contains 4 query heads. The results for both methods — measured across different NSA hyperparameters, sequence lengths, and tensor-parallel degrees — are summarized in the Table \ref{tab:h20-dist-attn} and \ref{tab:h200-dist-attn}. Compared to NSA, FSA achieves an average speedup of 1.16x on H20 GPUs and 1.17x on H200 GPUs. 

\subsection{End-to-end Distributed Inference Evaluation}

Following the same configuration as Figure \ref{fig:prefill}, we evaluate the distributed inference performance of the Llama-3-8B model on H20 and H200 GPUs. The results for NSA and FSA, measured under varying NSA hyperparameters, sequence lengths, and tensor-parallel degrees, are presented in the Table \ref{tab:h20-dist-e2e} and \ref{tab:h200-dist-e2e}. Compared to NSA, FSA achieves an average speedup of 1.13x on H20 GPUs and 1.11x on H200 GPUs. 

\begin{table}[h]
\centering
\caption{\ryan{End-to-end distributed inference latency on H20 GPU.}}
\label{tab:h20-dist-e2e}
\ryan{
\begin{tabular}{c c c c c c c}
\hline
($B_K$, $T$) & Seqlen (K) & Framework & TP=1 (s) & TP=2 (s) & TP=4 (s) & TP=8 (s) \\
\hline
(64, 16)  & 32 & FSA & 5.28 & 2.88 & 1.66 & 1.04 \\
(64, 16)  & 32 & NSA & 6.00 & 3.22 & 1.84 & 1.06 \\
(64, 16)  & 64 & FSA & 11.14 & 7.08 & 3.92 & 2.60 \\
(64, 16)  & 64 & NSA & 12.04 & 7.86 & 4.18 & 2.66 \\
(128, 8)  & 32 & FSA & 5.14 & 2.78 & 1.58 & 1.00 \\
(128, 8)  & 32 & NSA & 6.40 & 3.32 & 1.80 & 1.10 \\
(128, 8)  & 64 & FSA & 12.00 & 6.38 & 3.60 & 2.16 \\
(128, 8)  & 64 & NSA & 13.72 & 7.10 & 4.00 & 2.12 \\
\hline
\end{tabular}
}
\end{table}

\begin{table}[h]
\centering
\caption{\ryan{End-to-end distributed inference latency on H200 GPU.}}
\label{tab:h200-dist-e2e}
\ryan{
\begin{tabular}{c c c c c c c}
\hline
($B_K$, $T$) & Seqlen (K) & Framework & TP=1 (s) & TP=2 (s) & TP=4 (s) & TP=8 (s) \\
\hline
(64, 16) & 32 & FSA & 1.95 & 1.12 & 0.70 & 0.46 \\
(64, 16) & 32 & NSA & 1.97 & 1.22 & 0.70 & 0.49 \\
(64, 16) & 64 & FSA & 4.51 & 2.83 & 1.48 & 1.09 \\
(64, 16) & 64 & NSA & 4.61 & 2.81 & 1.51 & 1.16 \\
(128, 8) & 32 & FSA & 1.82 & 1.04 & 0.64 & 0.45 \\
(128, 8) & 32 & NSA & 2.37 & 1.33 & 0.78 & 0.53 \\
(128, 8) & 64 & FSA & 3.61 & 2.13 & 1.34 & 0.98 \\
(128, 8) & 64 & NSA & 4.62 & 2.63 & 1.61 & 1.15 \\
\hline
\end{tabular}
}
\end{table}
}

\end{document}